\documentclass[usenatbib,onecolumn]{mn2e}
\usepackage[authoryear]{natbib}
\usepackage{graphicx}
\newcommand\tempo{\textsc{tempo}}
\newcommand\tempoII{\textsc{tempo2}}

\newcommand\TempoII{\textsc{Tempo2}}
\newcommand\param[1]{\texttt{#1}}
\renewcommand\vec[1]{\ensuremath{\bmath{#1}}}
\newcommand\unitvec[1]{\ensuremath{\bmath{\hat{#1}}}}
\newcommand\mat[1]{\ensuremath{\mathbfss{#1}}}
\newcommand\cross{\ensuremath{\bmath{\times}}}
\newcommand\dotprod{\ensuremath{\bmath{\cdot}}}

\title[Tempo2: Timing model and precision]{\TempoII, a new pulsar timing package. II: The timing model and precision estimates}
\author[Edwards, Hobbs \& Manchester]{R. T. Edwards\thanks{Email: Russell.Edwards@csiro.au}, 
G. B. Hobbs, R. N. Manchester \\
Australia Telescope National Facility, CSIRO, P.O. Box 76, 
Epping NSW 1710 Australia}
\pubyear{2005}

\begin{document}

\maketitle

\begin{abstract}
\TempoII\ is a new software package for the analysis of pulsar
pulse times of arrival. In this paper we describe in detail
the timing model used by \tempoII, and discuss limitations on the
attainable precision. In addition to the intrinsic slow-down behaviour
of the pulsar, \tempoII\ accounts for the effects of a binary
orbital motion, the secular motion of the pulsar or binary system, interstellar, Solar system and ionospheric dispersion,
observatory motion (including Earth rotation, precession, nutation,
polar motion and orbital motion), tropospheric propagation delay,
and gravitational time dilation due to binary companions and Solar system
bodies. We believe the timing model is accurate in its
description of predictable systematic timing effects to better than
one nanosecond, except in the case of relativistic binary systems
where further theoretical development is needed. The largest remaining
sources of potential error are measurement error, interstellar
scattering, Solar system ephemeris errors, atomic clock instability
and gravitational waves.
\end{abstract}


\section{Introduction}
The core of any pulsar timing software package is the mathematical
model for the times of arrival (TOAs) of pulses at the observatory. As
the pulsar population grows more diverse and observing systems
increase in sensitivity, the accuracy demanded of timing models grows
accordingly.  For many years the standard choice for pulsar timing was
the \tempo\footnote{http://www.atnf.csiro.au/research/pulsar/tempo/}
software package.  Accurate at the 100-ns level, the timing model of
\tempo\ was adequate for numerous important discoveries
(e.g. \citealt{tfm79,tw89,wf92,ktr94,vbb+01}).  However, motivated
primarily by the possibility of directly detecting gravitational
radiation through its effect on pulse arrival times, current
observing campaigns are setting highly ambitious goals in terms of
data precision and volume. These projects aim to obtain of the order
of $\sim10^4$ independent TOA measurements with a typical
root-mean-square (RMS) uncertainty of 100~ns (e.g. \citealt{jhkm05}),
dictating a maximum allowable systematic error of 1~ns.

Motivated by the need for improved accuracy as well as greater
flexibility in fitting and analysis procedures, we have developed a
new pulsar timing package called \tempoII. The architecture,
capabilities and usage of \tempoII\ are described by \citet{hem06}
(hereafter Paper I).  The purpose of the present work is to describe
the timing model in detail, and provide estimates of the accuracy of
each component of the model.  Section \ref{sec:model} begins with an overview
of the model and 
definitions of the various relativistic coordinate frames and proper times
(Section \ref{sec:frames}). This is followed by an expansion of the
complete geometric propagation delay accurate to better than 1~ns
(Section \ref{sec:geometric}). The remainder of Section \ref{sec:model}
discusses all oWther delay terms in the timing formula, and provides
a detailed description of how each contribution is evaluated according
to the parameters of the model and other information.
Section \ref{sec:accuracy} gives a detailed account of the accuracy with
which each term is computed, and discusses possible avenues for further
improvement of the timing model. Finally, for the convenience of
comparison with the commonly used \tempo\ software package, 
Section \ref{sec:differences} details the differences between \tempo\ and
\tempoII. Tabular summaries of timing model parameters and variables
are provided in the Appendix.

\section{Timing Model}
\label{sec:model}
\subsection{Overview}
Pulsar timing consists of the measurement of the pulse times of
arrival (TOAs) at the observatory and the fitting of these TOAs to a
model.  The timing model relates the measured TOA to the time of
emission at the pulsar, from which a ``pulse phase of emission''
is computed via a model of the intrinsic variations in the pulse
period. If the timing model is correct, in the absence of measurement
noise all calculated pulse phases will be integer (when expressed in
units of cycles or turns). In practise, the parameters of the model
are not known a priori, and measurement noise is present, so that the
deviations of these pulse phases from the nearest integer (the
``residuals'') are finite.  Once the timing model parameters are
refined to optimise some statistic of the residuals -- in \tempoII\
this is the root-mean-square or $\chi^2$ (see Paper I) -- and
assuming there are no shortcomings in the structure of the timing
model, the residual deviations are due only to measurement noise.

The focus of this paper is the timing model, which relates a measured
TOA to a time of emission, and from there computes a corresponding
pulse phase of emission. The latter step is described in Section
\ref{sec:phase}.  The former part comprises the bulk of the complexity
of the model, and it too can be helpfully broken into two main
parts. Firstly, because that part of the model relates a proper time
of arrival at the observatory to a proper time of emission at the
pulsar, a general relativistic frame transformation is necessary. This
is performed in a sequence of manageable steps, outlined in Section
\ref{sec:frames}. Secondly, the light travel time must be accounted
for. This is derived in Section \ref{sec:geometric} under the
assumption of Euclidean space-time and zero refraction or dispersion.
The remaining parts of Section \ref{sec:model} derive in full
mathematical detail the various parts of the frame transformation and
light travel time, including dispersion, refraction and general
relativistic effects along the propagation path.

\subsection{Coordinate frames and time variables}
\label{sec:frames}
The relativistic frame transformation between observatory proper time
and pulsar proper time is computed in several steps. This section
outlines the steps, gives the names of the various reference frames
and introduces the notation use to refer to time variables in each
frame. The time variable for each frame may appear subscripted with
the letter ``a'' or ``e'', where ``a'' denotes a time of arrival and
``e'' a time of emission, at the origin of the frame denoted by the
superscript. A time variable may also appear without a subscript,
meaning that it is an an arbitrary coordinate time variable. As such,
time variables without subscripts are related by coordinate
transformations, whereas subscripted time variables additionally
include propagation delays. Spatial 3-vectors are defined in terms of
a right-handed Cartesian system.

The pulsar timing procedure begins with a measured arrival time, which
after correcting for imperfections in the observatory clock yields a
time of arrival $t^{\rm obs}_{\rm a}$. The clock correction process in
\tempoII\ operates through interpolation of tables of measured offsets
between pairs of clocks and their variation with time (see Paper I),
and includes offsets relating to the fact that observatory clocks
usually approximate Coordinated Universal Time (UTC\footnote{Due to
copyediting errors, in Paper I the acronyms UTC, TCG, TCB, and TDB
were inadvertanty referred to as UCT, CGT, BCT and BDT.}) rather than
Terrestrial Time (TT).  The corrected arrival time is measured in the
TT time scale, which is defined to differ from coordinate time at the
geocentre by a constant rate. It should be noted then that times
referred to TT are only approximately proper times of the
observatory. This transformation to Geocentric Coordinate Time scale
(TCG) is done in the same step as transformation to Barycentric
Coordinate Time (TCB), so no variable is assigned to this time
scale. The spatial part of the terrestrial frame is the International
Terrestrial Reference System (ITRS), while the geocentric counterpart
is the Geocentric Celestial Reference System (GCRS).

Coordinate times $t^{\rm SSB}$ in the frame of the Solar system
barycentre (SSB) are obtained by application of the gravitational
redshift and special relativistic time dilation integral (``Einstein
delay'') described in Section \ref{sec:einstein}. Coordinate time
(TCB) in this frame represents the proper time that would be measured
by an observer located at the Solar system barycentre were the
gravitational field of the sun and planets not present. The spatial
part of the barycentric frame is the Barycentric Celestial Reference
System (BCRS), while directions in this frame define the International
Celestial Reference System (ICRS). The origin of the BCRS is the SSB,
hence $t^{\rm SSB}_{\rm a}$ refers to the arrival time of the pulse at
the barycentre.

For further information on the definitions of TT, TCG, TCB, the ITRS,
GCRS, BCRS and ICRS, see the Resolutions of the XXIIIrd and XXIVth
General Assembly of the International Astronomical Union (IAU;
\citealt{and01,ric01}) and the 2003 Conventions of the International Earth
Rotation and Reference Systems Service (IERS; \citealt{mp04}).

Tracing the propagation backward toward the source, the next reference
frame is that which is co-moving with the pulsar, or if it is a member
of a binary system, the binary barycentre (BB). Coordinates in this frame
are defined such that the coordinate time $t^{\rm BB}$ would be equal
to the proper time of an observer at the origin of the frame, were the
gravitational field of the pulsar and its companion not present. The
transformation of the arrival time to this frame is discussed in
Section \ref{sec:doppler}. The orientation of this frame is chosen
such that the Lorentz transformation relating it to the SSB frame is
rotation-free, so that to zeroth order the corresponding basis vectors
are parallel. The spatial origin of this frame is the BB. The time origin
is chosen such that $t^{\rm BB}$ has zero offset from $t^{\rm SSB}$
at $t^{\rm BB}=t^{\rm SSB}=t_{\rm pos}$, where the latter is an epoch
of position set by the user.

The time scale for measurements is the proper time $t^{\rm psr}$
measured at the pulsar, or more precisely at its centre, were its
gravitational field not present. For isolated pulsars 
$t^{\rm psr}=t^{\rm BB}$. For binary pulsars, $t^{\rm psr}-t^{\rm BB}$ is
the binary Einstein delay (Section \ref{sec:binaryeinstein}). As with
the SSB--BB case, the BB--pulsar transformation is rotation-free. The
spatial origin of this frame is the centre of gravitational field of
the pulsar, so that $t^{\rm psr}_{\rm e}$ is the equivalent time of
emission from the centre of the star. This differs from the actual time
of emission by (to first order) the light travel time to the true point
of emission, which is assumed to be stable on the time scale of interest.

\subsection{Geometric propagation delay}
\label{sec:geometric}
Although it is convenient to break the propagation delay into separate
contributions and treat them in the order in which they occur along
the photon path, some terms in the model depend jointly upon the
motion of the observatory and the motion of the pulsar (relative to
the SSB). For this reason, in this section we derive the complete
geometric propagation delay, that is the time of propagation
neglecting dispersion, refraction and relativistic delays. The
resultant terms are then assigned to different parts of the propagation
path and treated in the relevant sections below.

The geometric propagation delay is simply the Euclidean distance from the
observatory to the pulsar, divided by the vacuum speed of light,
$c$. The displacement vector from the observatory to the pulsar is the
sum of the barycentric position of the observatory ($\vec{r}$), the
barycentric position of the pulsar (or BB)
($\vec{R_0}$) at a given epoch ($t_{\rm pos}$), the position of the
pulsar with respect to the BB ($\vec{b}$, zero for
isolated pulsars), and the displacement ($\vec{k}$) of the binary
barycentre (or the pulsar itself, if isolated) in the time elapsed
since the epoch $t_{\rm pos}$, owing to the initial velocity of the system
(relative to the SSB) and acceleration in the Galactic or cluster
gravitational field:
\begin{equation}
\vec{R} = \vec{R_0}+\vec{b}+\vec{k}-\vec{r}. \label{eq:displacement}
\end{equation}
\noindent Squaring equation (\ref{eq:displacement}) and expanding the square
root to third order,
\begin{eqnarray}
|\vec{R}| &=& \left[|\vec{R_0}|^2 + |\vec{b}|^2 + |\vec{k}|^2 + |\vec{r}|^2
 + 2\left(\vec{R_0}\cdot\vec{b}+\vec{R_0}\cdot\vec{k}-\vec{R_0}\cdot\vec{r}
         +\vec{b}\cdot\vec{k}-\vec{b}\cdot\vec{r}-\vec{k}\cdot\vec{r}\right)
      \right]^{1/2}\\
 &=& |\vec{R_0}| + A 
      - \frac{A^2}{2|\vec{R_0}|} + \frac{A^3}{2|\vec{R_0}|^2} + \ldots,\;
{\rm where} \\
A &=& \frac{1}{2|\vec{R_0}|}\left[|\vec{b}|^2 + |\vec{k}|^2 + |\vec{r}|^2
 + 2\left(\vec{R_0}\cdot\vec{b}+\vec{R_0}\cdot\vec{k}-\vec{R_0}\cdot\vec{r}
         +\vec{b}\cdot\vec{k}-\vec{b}\cdot\vec{r}-\vec{k}\cdot\vec{r}\right)
    \right].
\end{eqnarray}
\noindent Neglecting terms of order $|\vec{R_0}|^{-3}$ and denoting
the initially radial and transverse components of each vector using
subscripts, i.e. $a_\parallel = \vec{a}\cdot \vec{R_0}$,
$\vec{a_\perp} = \vec{a} - a_\parallel \vec{R_0}/|\vec{R_0}|$ and
hence, $\vec{a_\perp}\cdot \vec{b_\perp} =
(\vec{a}\cross\vec{R_0})\cdot(\vec{b}\cross\vec{R_0})$, the following
relation is found:
\begin{eqnarray}
|\vec{R}| &=& |\vec{R_0}|+k_\parallel-r_\parallel+b_\parallel \nonumber\\
&\;& + \frac{1}{|\vec{R_0}|}\left(
\frac{|\vec{k_\perp}|^2}{2} - \vec{k_\perp}\cdot\vec{r_\perp} + \vec{k_\perp}\cdot\vec{b_\perp}
+\frac{|\vec{r_\perp}|^2}{2} - \vec{r_\perp}\cdot\vec{b_\perp}  + \frac{|\vec{b_\perp}|^2}{2}
\right)
\left(1 - \frac{k_\parallel}{|\vec{R_0}|} + \frac{r_\parallel}{|\vec{R_0}|}
 - \frac{b_\parallel}{|\vec{R_0}|}\right) + O(|\vec{R_0}|^{-3}).
\label{eq:geometric}
\end{eqnarray}

\noindent The first four terms in equation (\ref{eq:geometric}) are
the initial SSB--BB distance (Section \ref{sec:vpd}), the secular
displacement in the initially radial direction (Section
\ref{sec:vpd}), the projection of the observatory--SSB vector on the
initial line of sight (Section \ref{sec:roemer}), and the projection
of the binary motion on the initial line of sight (Section
\ref{sec:binaryroemer}). The terms in the first pair of parentheses
correspond to the Shklovskii effect (\ref{sec:vpd}), annual proper
motion (\ref{sec:roemer}), secular changes in the apparent orbital
viewing geometry (\ref{sec:binarykop}), annual parallax
(\ref{sec:roemer}), annual-orbital parallax (\ref{sec:binarykop}) and
orbital parallax (\ref {sec:binarykop}). The second, third and fourth
terms in the second pair of parentheses represent variations in the
effects just discussed due to radial motion of the pulsar or binary
system, the motion of the Earth around the SSB, and the motion of the
pulsar about the BB. Although these terms give rise to an additional
eighteen contributions to the timing formula, to reach the stated
accuracy goal of 1~ns, for a 20-yr observing campaign on any of the
presently known millisecond pulsars only three of them need be
considered: the secular change (Section \ref{sec:vpd}) and annual
modulation (Section \ref{sec:roemer}) of the Shklovskii effect and the
secular change of the annual proper motion (Section \ref{sec:roemer}).

The displacement due to secular motion, $\vec{k}$ may be broken into
first and second derivatives:
\begin{equation}
\vec{k} = \vec{\mu}|\vec{\rm R_0}|\left(t^{\rm BB}-t_{\rm pos}\right)
 + \frac{\vec{a}}{2}\left(t^{\rm BB}-t_{\rm pos}\right)^2 , \label{eq:k} 
\end{equation}
\noindent where $\vec{\mu}$ is the velocity divided by the distance,
or in essence a three-dimensional proper motion, and $\vec{a}$ is an
acceleration vector accounting for Galactic differential rotation and
gravitational acceleration. The acceleration is of order $\sim
10^{-11}$~m~s$^{-2}$ \citep{bb96}, contributing only $\sim 500$~km to
$\vec{k}$ at either end of a 20-year observation campaign centred on
$t_{\rm pos}$. Equation (\ref{eq:geometric}) can be expanded in terms
of this parameterisation of $k$, giving rise to an additional 27
terms. We have estimated the magnitude of these terms for all pulsars
in the ATNF
catalogue\footnote{http://www.atnf.csiro.au/research/pulsar/psrcat/}
\citep{mhth05},
including an allowance for greater acceleration for pulsars in
globular clusters ($a \sim 10^{-8}$~m~s$^{-2}$; \citealt{fcl+01}).
Only two of the extra terms involving $\vec{a}$ exceed 1~ns for a
20-year observing campaign, the first contributing to $k_\parallel$
and the second contributing to the Shklovskii term of equation
(\ref{eq:geometric}). For all other terms of equation
(\ref{eq:geometric}) involving $\vec{k}$ the acceleration is
neglected.

\subsection{Top-level timing formula}
Having described the coordinate frames used and derived the basic
Euclidean propagation length, it remains to specify the details of
the calculation of the various delay terms associated with frame
transformations, the vacuum propagation delay, and corrections due
to non-unity refractive indices and space-time curvature along the
propagation path. For convenience the timing formula is divided into
three main contributions, which will be further broken down in the
sections that follow:
\begin{equation}
t^{\rm psr}_{\rm e} = t^{\rm obs}_{\rm a} - \Delta_\odot - \Delta_{\rm IS} - \Delta_{\rm B} \label{eq:top}.
\end{equation}

$\Delta_\odot$ (Section \ref{sec:delta_SSB}) includes the coordinate
transformation to the SSB frame, vacuum propagation delays associated
with the Earth's orbital motion and the spin, precession and nutation
of the Earth (on which the observatory is located) and excess delays
owing to the passage of the signal through the Earth's atmosphere and
the Solar system. This term relates the measured time of arrival to
the TCB arrival time at the SSB: $t^{\rm SSB}_{\rm a}= t^{\rm
obs}_{\rm a} - \Delta_\odot$.

$\Delta_{\rm
IS}$ (\ref{sec:ISM}) includes the transformation to the binary
barycentre frame, vacuum propagation delays due to the secular motion
of the system, and excess propagation delays due to passage of the
signal through the interstellar medium. This term relates the
TCB time of arrival at the SSB to the BB coordinate time of arrival at
the BB: $t^{\rm BB}_{\rm a} = t^{\rm SSB}_{\rm a}-\Delta_{\rm IS}$.

$\Delta_{\rm B}$ includes the transformation to the pulsar frame,
vacuum delays due to the binary orbital motion, and excess delays due
to passage of the signal through the gravitational field of the
companion.  This term relates the BB coordinate time of arrival at the
BB to the pulsar proper time of emission: 
$t^{\rm psr}_{\rm e}=t^{\rm BB}_{\rm a} - \Delta_{\rm B}$.

Note that since some of the
geometric delay terms involve two of the three components (Solar
system, secular or binary) of the relative motion of the observatory
and the pulsar, the choice of which of these terms are assigned to
each of the three top-level terms is somewhat arbitrary. See Section
\ref{sec:geometric} for a detailed listing of where each term is
discussed. 

Once the time of emission is obtained, it remains to compute the phase
of the pulse train at that time. This relationship, $\phi(t^{\rm
psr}_{\rm e})$, is discussed in Section \ref{sec:phase}.  For the
correct set of model parameters and in the absence of measurement
noise, the phase for every computed time of emission is an integer
number of cycles (turns). The fractional part of $\phi(t)$ is the
timing residual, which is more often expressed in time units through
division by the pulse frequency.  

\subsection{Forming barycentric arrival times}
\label{sec:delta_SSB}
The term denoted $\Delta_\odot$ in equation (\ref{eq:top}) includes
all effects pertinent to the transformation of the observed TOA
($t_{\rm a}^{\rm obs}$) to an equivalent time of arrival for the same
pulse wave-front at the Solar system barycentre.  This in turn can be
broken into several steps. These are: atmospheric delays, vacuum
retardation due to observatory motion (``Roemer delay'' and parallax),
excess propagation delay due to dispersion, the effects of
relativistic frame transformation (``Einstein delay''), and the excess
delay experienced by rays as they pass through the gravitational
potential of Solar system bodies (``Shapiro delay''). These terms are
denoted as follows:
\begin{equation}
\Delta_\odot =  \Delta_{\rm A} + 
   \Delta_{\rm R\odot} + \Delta_{\rm p} + \Delta_{{\rm D}\odot} + \Delta_{{\rm E}\odot}  
  + \Delta_{\rm S\odot}
\label{eq:deltaSSB}
\end{equation}

\subsubsection{Einstein delay}
\label{sec:einstein}
  The space-time coordinates of pulse arrival events are specified in
  the coordinate frame of the observatory, in which pulse timing
  anomalies may manifest due to spin and orbital acceleration and
  variations in gravitational potential. To avoid these effects, pulse
  arrival events are transformed to the quasi-inertial frame of the
  Solar system barycentre. The two frames are related by a
  relativistic four-dimensional space-time transformation.  The
  spatial part of the event is the displacement of the observatory
  from the geocentre at the instant of reception, which is altered by
  a negligible amount due to special and general relativistic length
  contraction. The effects of relativistic time dilation, on the other
  hand, cannot be neglected. \tempoII\ uses the numerical time dilation
  results of \citet{if99}, who used the DE405 Solar system ephemeris
  \citep{sta98b}
  to compute the time dilation integral:
\begin{equation}
\Delta_{{\rm E}\odot-\oplus} = \frac{1}{c^2}\int_{t_0}^t 
      \left(U_\oplus + \frac{v_\oplus^2}{2}
       +\Delta L_C^{\rm (PN)} + \Delta L_C^{\rm (A)} \right) {\rm d}t,
\label{eq:timedilation}
\end{equation}
\noindent where $U_\oplus$ is the gravitational potential at the geocentre
due to all Solar system bodies except the Earth, and $v_\oplus$ is the
velocity of the geocentre relative to the Solar system barycentre.
The first two terms describe, to order $1/c^2$, the gravitational
redshift and special relativistic time dilation respectively. These
contribute in roughly 2:1 ratio ($\frac{1}{2}U_\oplus \simeq v_\oplus^2 \simeq 10^{-8}
c^2$), for a mean drift of $\sim 1.5\times 10^{-8}$, or $\sim 0.5$
s yr$^{-1}$. The main source of time variation in the integrand is due to
the orbital acceleration of the Earth, which via conservation of
energy causes approximately equal changes (of the same sign) in the
first two terms, amounting to $\sim 3 \times 10^{-10}$ in rate
amplitude, or $\sim 2$~ms in integrated delay amplitude. The remaining
terms apply a correction for higher-order relativistic terms ($\Delta
L_C^{\rm (PN)} = 1.097 \times 10^{-16}$; \citealt{fuk95}) and
asteroids ($\Delta L_C^{\rm (A)} = 5 \times 10^{-18}$;
\citealt{fuk95}), made in mean rate only.

The time dilation integral of equation (\ref{eq:timedilation}) relates
the coordinate times of clocks at the geocentre and the Solar system
barycentre. For observers located on the surface of the Earth,
corrections must be made for the differential time dilation and
gravitational redshift between the observatory and the geocentre,
yielding:
\begin{equation}
\Delta_{E\odot} = \Delta_{{\rm E}\odot-\oplus} 
  + \frac{\vec{s}\dotprod \vec{\dot{r}_\oplus} + W_0t_{\rm a}^{\rm obs}}{c^2},
\label{eq:einstein}
\end{equation}
\noindent where $\vec{s}$ is a vector from the geocentre to the observatory,
$\vec{\dot{r}_\oplus}$ is the velocity of the geocentre with respect
to the barycentre, and $W_0 = 6.969290134\times 10^{-10} c^2$
approximates the gravitational plus spin potential of the Earth at the
geoid. (See Section \ref{sec:roemer} for the calculation of $\vec{s}$
and $\vec{\dot{r}_\oplus}$.) Note that, while the value of the
potential varies geographically and temporally, the most recent
definition of TT \citep{ric01}, the terrestrial time scale to which
\tempoII\ refers pulse time of arrival measurements, takes the exact
value of $W_0/c^2$ quoted above as definitive of the rate difference
between TT and SI coordinate time at the geocentre. Given a common
epoch of $t_0=43144.0003725$ (Modified Julian Date),
equations \ref{eq:timedilation} and \ref{eq:einstein} relate measured TT to
the IAU recommended coordinate timescales TCG (geocentric) and TCB
(barycentric).

It should be noted that the time ephemeris of \citet{if99} in fact
takes its time argument, $T_{\rm eph}$, in a reference frame that is a
linear transformation of TCB, making the computation of the Einstein
delay strictly an inverse problem. In practise, since the argument
time scale has zero mean rate difference to TT, the latter can be
substituted with negligible error in the result \citep{if99}. In
addition, the evaluation of equation (\ref{eq:einstein}) requires consulting
the Solar system ephemeris (Section \ref{sec:roemer}) which also takes
$T_{\rm eph}$ times as its argument. Since
$T_{\rm eph}$ is computed from TCB (equation \ref{eq:einstein}),
a circular dependency problem arises. This is solved to sufficient
accuracy by two passes through an iterative refinement process.

\subsubsection{Atmospheric Delays}
\label{sec:atm}
The group velocity of radio waves in the atmosphere differs from
the vacuum speed of light. Refractivity is induced both by the
ionised fraction of the atmosphere (mainly in the ionosphere) and
the neutral fraction (mainly in the troposphere). 

The tropospheric propagation delay be separated into the so-called
``hydrostatic'' and ``wet'' components. In Very Long Baseline
Interferometry (VLBI) and Global Positioning System (GPS)
applications, these components are typically modelled as the product of
the delay induced at the zenith, and a so-called ``mapping function'',
which specifies the ratio between the zenith delay and the delay in a
given direction.  For a planar atmosphere, the mapping function is
simply given by $\csc\Theta$ where $\Theta$ is the source
elevation angle.  More advanced mapping functions take into account
the curvature of the atmosphere, assuming azimuthal symmetry. The
effect of azimuthal asymmetry is typically less than a nanosecond even
at very low elevation angles (e.g. \citealt{mac95}), and need not be
considered for the present purpose.

The hydrostatic component contributes approximately 90\% 
of the total delay, and may be computed a priori assuming a
given mixture of gases in hydrostatic equilibrium. \tempoII\
uses the formula of \citet{dhs+85}, re-written as a time delay
\begin{equation}
\Delta_{\rm hz} = 
\frac{\left(\frac{P}{43.921\;\mathrm{kPa}}\right)}
 {c\left[1-0.00266\cos 2\varphi - 
\left(\frac{3.6\times 10^{6}\;\mathrm{m}}{H}\right)\right]}
 ,
\label{eq:hydrostatic_zenith_delay}
\end{equation}
where $\Delta_{\rm hz}$ is the hydrostatic zenith delay, $P$ is the
surface atmospheric pressure, $\varphi$ is the geodetic latitude of the
site, and $H$ is its height above the geoid.  In combination with the
mapping function, this leads to a timing term of amplitude $\sim 7.7$
ns $\cdot \csc \Theta$, mostly on a diurnal timescale.  If
atmospheric pressure data are unavailable, \tempoII\ uses a canonical
value of one standard atmosphere (101.325 kPa). \tempoII\ uses the
Niell Mapping Function (NMF; \citealt{nie96}), which is of comparable
accuracy to other published mapping functions but does not require
meteorological data: it depends only on the source elevation.

The wet component of the tropospheric propagation delay is highly
variable and cannot be predicted accurately. Fortunately, it is small.
The zenith wet delay (ZWD; $\Delta_{\rm wz}$) may be measured using
appropriate analyses of radiosonde, water vapour radiometer, GPS or
VLBI observations (e.g. \citealt{ncs01}), of which only GPS is likely
to be obtainable on a routine basis at most radio observatories. If
such measurements are available, \tempoII\ can make use of them in
conjunction with the NMF to account for the effects of tropospheric
water vapour on pulse times of arrival.  The ZWD could conceivably be
included as a free parameter in the pulsar timing model, but because
the effect is small and varies from day to day, this would not be
possible with the sensitivity of presently attainable observing
systems. For this reason, if no tabulated ZWD information is
available, the effect is neglected.

Neglecting dispersion (below), the total atmospheric delay is written:
\begin{equation}
\Delta_{\rm A} = m_h(\Theta)\Delta_{\rm hz}(P) + m_w(\Theta)\Delta_{\rm wz}(P) ,
\end{equation}
\noindent where $m_h(\Theta)$ and $m_w(\Theta)$ are the Niell mapping
functions for the hydrostatic and wet components. Values for 
$P$ and $\Delta_{\rm wz}$ are to be provided as input data to \tempoII,
or default values of $P=101.325$~kPa and $\Delta_{\rm wz}=0$ are used.

The passage of the signal through the ionosphere induces a dispersive
delay that varies strongly with the Solar activity cycle and also on
on seasonal, diurnal and shorter timescales. The integrated column
density of electrons (``total electron content''; TEC) in the
ionosphere typically lies in the range 5--100 TECU (1 TECU
$=10^{16}$~m$^{-2} \simeq 3.2 \times 10^{-7}$~cm$^{-3}$~pc),
corresponding to variable propagation delays of the order of
7$-$130~ns~$\cdot(f/1\;\mathrm{GHz})^{-2}$, where $f$ is the
observing frequency. Models of the ionosphere are available
(e.g. \citealt{sch99,bil01}), however the predicted TEC is accurate to
only $\sim$ 3--20 TECU \citep{mkk03}. In any case, ionospheric
dispersion is inseparable from and smaller than uncertainties in
interplanetary (Section \ref{sec:ssdm}) and interstellar
(Section \ref{sec:isdm}) dispersion, necessitating the inclusion of a
fittable, fully time-variable dispersion parameter
(Section \ref{sec:isdm}).

\subsubsection{Roemer delay and parallax}
\label{sec:roemer}
The  Roemer delay is the simple vacuum delay between the
arrival of the pulse at the observatory and the Solar system
barycentre, not including effects related to the binary motion of, or
finite distance to, the pulsar:
\begin{equation}
\Delta_{\rm R\odot} = -\frac{\vec{r} \dotprod \unitvec{R}_{\rm BB}}{c},
\label{eq:roemer}
\end{equation}
\noindent where $\unitvec{R}_{\rm BB} \equiv \vec{R}_{\rm
BB}/|\vec{R}_{\rm BB}|$ is a unit vector in the direction of the
binary barycentre at the time of observation. \tempoII\ performs this
calculation in the BCRS.

The vector $\unitvec{R}_{\rm BB}$ is constructed from the spherical
coordinate angles of the ICRS source direction, right ascension
($\alpha$) and declination ($\delta$), at time $t_{\rm pos}$, and the
Cartesian components of the proper motion of the source in the plane
of the sky ($\mu_\alpha$ and $\mu_\delta$) and along the line of sight
($\mu_\parallel$), all of which are fittable parameters in \tempoII:
\begin{eqnarray}
\unitvec{R}_{\rm BB} &=& \unitvec{R_0} 
+ \vec{\mu_\perp}\left(t^{\rm BB}_{\rm a}-t_{\rm pos}\right)
   - \left(\frac{1}{2}|\vec{\mu_\perp}|^2 \unitvec{R_0} 
  + \mu_\parallel\vec{\mu_\perp}\right)
\left(t^{\rm BB}_{\rm a}-t_{\rm pos}\right)^2,  
{\rm where} \label{eq:sourcevec}\\
\unitvec{R_0} &=& \left(\begin{array}{c} 
  \cos\alpha\cos\delta \\
  \sin\alpha\cos\delta \\
  \sin\delta
  \end{array}\right) \\
\vec{\mu_\perp}  &=& \mu_\alpha\unitvec{\alpha} + \mu_\delta\unitvec{\delta},\\
\unitvec{\alpha} &=& \left(\begin{array}{c}
-\sin\alpha \\ \cos\alpha \\ 0 \end{array}\right),\; {\rm and}\\
\unitvec{\delta} &=& \left(\begin{array}{c}
-\cos\alpha\sin\delta \\ -\sin\alpha\sin\delta \\ \cos\delta\end{array}\right).
\end{eqnarray}
\noindent Alternatively, ecliptic coordinates ($\lambda$, $\beta$,
$\mu_{\lambda}$, $\mu_{\beta}$) may be used, in which case:
\begin{eqnarray}
\unitvec{R_0} &=& \mat{E}\left(\begin{array}{c} 
  \cos\lambda\cos\beta \\
  \sin\lambda\cos\beta \\
  \sin\beta
  \end{array}\right),\;{\rm and} \\
\vec{\mu_\perp}  &=& \mu_\lambda\unitvec{\lambda} + \mu_\beta\unitvec{\beta},
\;{\rm where}\\
\mat{E} &=& \left(\begin{array}{ccc}
1 & 0 & 0\\
0 & \cos\epsilon_0 & -\sin\epsilon_0 \\
0 & \sin\epsilon_0 & \cos\epsilon_0 
\end{array}\right), \\
\unitvec{\lambda} &=& \mat{E}\left(\begin{array}{c}
-\sin\lambda \\ \cos\lambda \\ 0 \end{array}\right),\; {\rm and}\\
\unitvec{\beta} &=& \mat{E}\left(\begin{array}{c}
-\cos\lambda\sin\beta \\ -\sin\lambda\sin\beta \\ \cos\beta\end{array}\right).
\end{eqnarray}
Here $\epsilon_0 = 84381.40578$~arcsec \citep{hf04} is the mean obliquity
of the ecliptic at J2000.0.

The four terms that result from the expansion of equation
(\ref{eq:sourcevec}) derive from corresponding terms in the expansion
of equation (\ref{eq:geometric}), such that
\begin{equation}
\vec{r} \dotprod \unitvec{R}_{\rm BB}
 = r_\parallel + \frac{\vec{k_\perp}\cdot\vec{r_\perp}}{|\vec{R_0}|}
   - \frac{r_\parallel|\vec{k_\perp}|^2}{2|\vec{R_0}|^2}
   - \frac{k_\parallel \vec{k_\perp}\cdot\vec{r_\perp}}{|\vec{R_0}|^2}.
\label{eq:rdotRhat}
\end{equation}
\noindent Here we have assumed that $\vec{k} = \vec{\mu}|\vec{R_0}|\left(t^{\rm
BB}_{\rm a}-t_{\rm pos}\right)$, safely neglecting any acceleration or
higher order secular motion of the pulsar or BB (Section
\ref{sec:geometric}). In principle, $t^{\rm BB}_{\rm a}$ cannot be
computed without first knowing $\Delta_{\rm R\odot}$; this is resolved
by initially setting $t^{\rm BB}_{\rm a}$ to the TCB time of arrival
at the observatory, and iteratively refining $\Delta_{\odot}$ and
$\Delta_{\rm IS}$ until convergence.


Equation \ref{eq:rdotRhat} consists of four terms corresponding to the
expression in equation \ref{eq:sourcevec} of the instantaneous source
direction as the sum of four contributions. These are the initial
direction (yielding the first term of equation \ref{eq:rdotRhat}), the
proper motion (second term), a correction to maintain unit length
after addition of the proper motion (third term) and the acceleration
or deceleration of the proper motion owing to the fact that over time,
the proper motion moves the source vector, causing part of the
initially radial velocity to acquire a transverse component (fourth
term). The third term can also be interpreted as the
proper-motion-induced second-order reduction in the projection of the
source direction on the radial component of $\vec{r}$ that accompanies
the first-order increase in its projection upon the transverse
component of $\vec{r}$. Another alternative interpretation of the
third term is that it corrects the Shklovskii term (Section
\ref{sec:vpd}) for the distance variations induced by the Earth's
orbital motion. The fourth term, first introduced by N.~ Wex (1997,
unpublished contribution to \tempo), provides one of two possible ways
of accessing the elusive radial velocity, the integrated first-order
contribution of which ($k_\parallel$) is otherwise inseparable from a
Doppler shift of the pulse and orbital frequencies and their
derivatives. The second method is described in Section \ref{sec:vpd}.


The first term of equation (\ref{eq:roemer}) describes the difference
in time of arrival of a pulse at the SSB compared to that at the point
obtained by projecting the SSB--observatory vector on the SSB--BB
vector. The arrival time at that point differs again from the arrival
time at the observatory due to the curvature of spherical
``wavefronts'' connecting photons emitted simultaneously from the
pulsar. This term is referred to as the ``parallax'' term, although it
differs somewhat from stellar parallax: positional astrometry measures
the three-dimensional orientation of the wavefront normal,
whereas pulsar timing measures the position of the wavefront in its
intersection with the ecliptic.  To first order the curvature is
proportional to the square of the lateral displacement of the
observatory from the SSB--BB vector, yielding an excess delay
corresponding to one of the terms of equation (\ref{eq:geometric}):
\begin{equation}
\Delta_{\rm p} = \frac{|\vec{r_\perp}|^2}{2cd_p}, 
\end{equation}
\noindent Here the ``parallax distance'', $d_p$ is substituted for
$|\vec{R_0}|$ to allow this effect to be separated from others
involving the source distance. The conventional ``parallax''
$\Pi \equiv 1\;{\rm AU}/d_p$, being
the angle subtended from the source by line of 1 AU in length at the
SSB, is a fittable parameter in \tempoII.  Its value is specified in
units of milliarcseconds, corresponding numerically to the reciprocal
of the distance in kiloparsecs.

In computing the Roemer and parallax delays, the vector $\vec{r}$ is
constructed in two steps, from the SSB to the geocentre and from the
geocentre to the observatory, i.e.  $\vec{r} = \vec{r_\oplus} +
\vec{s}$. The first part ($\vec{r_\oplus}$) is accomplished with the
aid of a numerical Solar system ephemeris.  The current default ephemeris is
DE405 \citep{sta98b}, which is aligned within a known uncertainty to
the ICRS, but uses a time and length scale slightly different than the
SI second and metre \citep{sta98c}. \tempoII\ uses the equations and
constants of \citet{if99} to appropriately transform the TCB site
arrival time (Section \ref{sec:einstein}) for input to the ephemeris
(using an offset and scale), and appropriately scales the output
vectors to SI units. It should be noted that the use of alternative
ephemerides that require a different transformation procedure will
introduce errors (see Section\ \ref{sec:accuracy:roemer}).

The second part of the observatory position vector, $\vec{s}$, is
obtained by transformation of the terrestrial observatory coordinates
into the GCRS (which to sufficient accuracy is equivalent to the
BCRS). This is accomplished using an ITRS site position vector
($\vec{s}_{\rm ITRS}$) supplied to \tempoII, transformed to the GCRS
using algorithms consistent with the IAU 2000 Resolutions. The
transformation proceeds as follows:
\begin{equation}
\vec{s} = \mat{Q}(t_{\rm a}^{\rm obs})\mat{R}(t_{\rm a}^{\rm obs})\mat{W}(t_{\rm a}^{\rm obs}) \vec{s}_{\rm ITRS},
\end{equation}
\noindent where the matrices $\mat{W}$, $\mat{R}$ and $\mat{Q}$
account for polar motion, Earth rotation, and the motion of the Earth
spin axis in the ICRS. The latter term in turn consists of a product
of a frame bias and the IAU2000B precession-nutation matrix
\citep{ml03}.  Details of the construction of these matrices are
available in the IERS Conventions \citep{mp04} and in the documentation of
the IAU Standards of Fundamental Astronomy (SOFA) software
library\footnote{http://www.iau-sofa.rl.ac.uk/} used by \tempoII\ to
compute them. Polar motion and Earth rotation both exhibit
unpredictable variations that must be corrected post facto using
measured parameters. \tempoII\ uses the ``C04'' series of Earth
orientation parameters, provided by the IERS, as input to the SOFA
routines.

For the purposes of transforming the observing frequency to the
barycentric frame (Section \ref{sec:isdm}), the time derivative of the
Roemer delay is needed:
\begin{equation}
\frac{{\rm d}\Delta_{R\odot}}{{\rm d}t} =
     \frac{\left(\vec{\dot{r}_\oplus}+\vec{\dot{s}}\right) \dotprod \unitvec{R_0}}{c},
\label{eq:roemer_rate}
\end{equation}
\noindent where dots denote the time derivative and the effects of
proper motion and parallax are negligible. The barycentric velocity
of the geocentre, $\vec{\dot{r}_\oplus}$, is provided by the Solar
system ephemeris. The barycentric velocity of the site is dominated
by the velocity imparted by Earth rotation. This is calculated using
\begin{equation}
\vec{\dot{s}}_{\rm ITRS} = \left(\omega_\oplus\mat{W}^{-1}\unitvec{z}\right)
  \cross \vec{s}_{\rm ITRS} .
\end{equation}
\noindent Here a unit North vector ($\unitvec{z}$) is transformed
using the inverse polar motion matrix and multiplied by the
instantaneous Earth angular rotation rate $\omega_\oplus$, to give the
angular velocity of the Earth in the ITRS.  The cross product of this
with the site radius vector gives the tangential velocity, which is
transformed to the GCRS in the same manner as $\vec{s}$. Of the
neglected terms in this approximation, the largest is the precessional
velocity, at a negligible $\sim 10^{-13}c$.

\subsubsection{Solar system dispersion}
\label{sec:ssdm}
Radio signals encounter significant dispersion in the interplanetary
medium, due to the electron content of the Solar wind. The electron
distribution follows a roughly $r^{-2}$ form consistent with spherical
expansion \citep{immh98}. Integrating along the line of sight 
 yields a dispersion measure (DM):
\begin{eqnarray}
{\rm DM}_{\odot} &=& \int_0^{\infty} n_{0} 
  \left(\frac{1\;{\rm AU}}{r(s)}\right)^2 {\rm d}s \\ 
   &=& n_{0}\left(1\;{\rm AU}\right)^2\frac{\rho}{|\vec{r}|\sin \rho},
\label{eq:ssdm}
\end{eqnarray}
\noindent where $|\vec{r}|$ is the heliocentric radius of the observatory,
$\rho$ is the pulsar-Sun-observatory angle, and $n_{0}$ is an overall
scale parameter, which can be specified in \tempoII\ parameter files
as \param{NE1AU}. (It should be noted that equation (\ref{eq:ssdm}) is an
approximation that ignores the effects on the dispersion relation due
to the bulk plasma velocity and its variation along the line of sight.
However, there are other much more significant sources of error as
discussed below.)  The default value for $n_{0}$ is 4~cm$^{-3}$,
consistent with recent measurements
\citep{immh98,ihmm01,sns+05} and significantly lower than the value of
9.961~cm$^{-3}$ used by \tempo. The computed dispersion measure is used
to calculate the delay due to interplanetary dispersion:
\begin{equation}
\Delta_{\rm D\odot} = \frac{{\rm DM}_{\odot}}
           {2.410 \times 10^{-16}\;{\rm cm}^{-3}\;{\rm pc}}
 \cdot \left(f^{\rm SSB}\right)^{-2}
\label{eq:ssdmdelay}
\end{equation}
\noindent where $f^{\rm SSB}$ is the observing frequency transformed to the barycentric
frame (see Section \ref{sec:isdm}).

This correction is of limited accuracy.  Recent studies of with the
{\it Ulysses} spacecraft have revealed extraordinary complexity in the
Solar wind. At Solar minimum, the electron density (scaled by $r^2$)
was found to show RMS temporal variations of 10--50\%, depending on
heliocentric latitude \citep{immh98}, while at Solar maximum the mean
electron density increased but was subject to very strong modulation
\citep{ihmm01}.  The default correction of equation (\ref{eq:ssdmdelay}) is
therefore uncertain within at least factor of two, corresponding to a
minimum of 130~ns of error at $f=1$~GHz for a source located at
an ecliptic pole, increasing to many microseconds for sources within a
few degrees of the ecliptic.

Errors in the predicted Solar system dispersion delay are of concern
not only because they add random noise to the timing residuals and
fitted parameters, but also because a portion of the error term will
exhibit annual periodicities that could corrupt the fitted pulsar
position, proper motion and parallax \citep{sns+05}. It is also
possible that other periodicities in this term could be misinterpreted
as arising due to a planetary companion to the pulsar
(\citealt{sfal97}, though see also \citealt{whk+00}). Previous
attempts to improve the model have focussed on adding fittable degrees
of freedom (e.g. \citealt{cbl+96,sns+05}), however the complex
behaviour seen in the {\it Ulysses} observations indicate that
the only way to adequately remove the effect is to measure it directly
using multi-frequency observations (see Section\ \ref{sec:isdm}).

\subsubsection{Shapiro delay}
\label{sec:shapiro}
The Shapiro delay accounts for the time delay caused by the passage of
the pulse through curved spacetime.  The total delay
obtained by summing over all the bodies in the Solar system \citep{bh86}:
  
\begin{equation}\label{eqn:shapiro}
  \Delta_{\rm S\odot} = -2\sum_j\frac{Gm_j}{c^3}\ln \left(\unitvec{R} \dotprod
  \vec{r_j} + |r_{j}|\right) + \Delta_{\rm S\odot 2}
\end{equation}
where  $\unitvec{R}$ is a unit vector
in the direction of the pulsar, $m_j$ is the mass of body $j$, $\vec{r_j}$
is a vector from body $j$ to the telescope, and $\Delta_{\rm S\odot 2}$
is a second-order correction discussed below. The first term may be re-written
in the following more convenient form:
\begin{equation}
  \Delta_{\rm S\odot} = -\sum_j\frac{2G m_j}{c^3}\ln |r_j|\left(1-\cos \psi_j\right) + \Delta_{\rm S\odot 2},
\end{equation}
\noindent where $\psi_j$ is the pulsar-telescope-object angle for
the $j$-th object. \textsc{Tempo2} includes the first-order Shapiro
delay for the Sun, Venus, Jupiter, Saturn, Uranus and Neptune.

For observations of sources very close to Solar system bodies,
higher-order effects may become important. The largest of these is the
geometrical excess path length due to gravitational light bending
\citep{rm83}, which amounts to $9.1$~ns for a ray with a trajectory
grazing the Solar limb\footnote{\citet{hel86} quotes a figure of
36~ns, referring to the calculation of \citet{rm83}. However, that
calculation incurred an extra pair of factors of two, one due to the
fact that the calculation referred to a round trip, and one due the
fact that the reflector was located 1 AU past the sun, rather than
being a very distant source.}. From the derivation of \citet{hel86},
assuming general relativity and re-parameterising in terms of $\psi$
for a very distant source, this term is given by
\begin{eqnarray}
  \Delta_{\rm S\odot 2} &=& \frac{4G^2m^2}{c^5|r|\tan\psi \sin\psi}\\
 &\simeq& \frac{4G^2m^2}{c^5|r|\psi^2}.
\end{eqnarray}
\noindent For pulsar timing purposes, only the geometric delay due to the Sun need be considered. 

\subsection{Propagation through interstellar space}
\label{sec:ISM}
The propagation time of the signal from the pulsar (or binary system barycentre)
to the SSB is conveniently broken into three contributions: the vacuum
propagation delay, i.e. the path length divided by the vacuum speed of
light, and the excess delay due to dispersion and other frequency-dependent
delays:
\begin{equation}
\Delta_{\rm IS} = \Delta_{\rm VP} + \Delta_{\rm ISD} + \Delta_{\rm
FDD} + \Delta_{\rm ES}.
\label{eq:ism}
\end{equation}
The fourth term accounts for the special relativistic time dilation between
the SSB reference frame and the binary reference frame. 

\subsubsection{Interstellar dispersion and other frequency-dependent delays}
\label{sec:isdm}
The interstellar dispersion delay enters the timing model under the
standard relation:
\begin{equation}
  \Delta_{\rm ISD} = \frac{D}{\left(f^{\rm SSB}\right)^2},
  \label{eq:dm}
\end{equation}
\noindent where $D$ is the so-called dispersion constant, and
$f^{\rm SSB}$ is the frequency of the radiation at the Solar system
barycentre. The barycentric frequency differs from the frequency at
the observatory, due to a simple Doppler shift ($\sim 10^{-4}$ in
magnitude) and a second-order relativistic correction ($\sim
10^{-8}$), plus the effects of gravitational redshift ($\sim 10^{-8}$).
The first term is simply the time derivative of the Roemer delay
(the ``Roemer rate''), while the sum of the latter two are is derivative of
the Einstein delay (the ``Einstein rate''). 

\begin{equation}
f^{\rm SSB}= \left(1+\frac{{\rm d}\Delta_{R\odot}}{{\rm d}t}+\frac{{\rm d}\Delta_{E\odot}}{{\rm d}t}\right) f .
\end{equation}

Although $D$ is the directly measurable quantity of equation
(\ref{eq:dm}), it is customary to speak in terms of the inferred
column density of electrons along the line of sight, i.e. the
dispersion measure, ${\rm DM}=k_{\rm D}D$. The value of $k_{\rm D}$
used by \tempoII\ (and \tempo) is $k_{\rm D}\equiv 2.410 \times
10^{-16}$~cm$^{-3}$~pc; see Paper I for further discussion. 

Many pulsars exhibit temporal variation in the interstellar dispersion
measure. \tempoII\ follows other software packages in modelling these variations
by means of a Taylor series:
\begin{equation}
{\rm DM} = \sum_{n\geq 0} \frac{{\rm DM}^{(n)}}{n!} 
       \left(t^{\rm SSB}_a-t_{\rm D}\right)^n,
\label{eq:dmtaylor}
\end{equation}
\noindent where ${\rm DM}^{(n)}$ is the (fittable) $n$-th derivative
of the dispersion measure and $t_{\rm D}$ is a user-specified epoch.

This approach is limited to modelling variations of low to moderate
complexity. At the level of accuracy demanded by current high precision
timing campaigns, significant unmodelled variations will appear
on short timescales, due to ionospheric (Section \ref{sec:atm}),
interplanetary (Section \ref{sec:ssdm}) and interstellar
electrons \citep{fc90}. An alternative and recommended course of action is
to use the ``stride fit'' feature of \tempoII\ (Paper I), to fit a 
different dispersion measure to every group of simultaneous 
(or contemporaneous) multi-frequency arrival time measurements.

\tempoII\ also allows for the fitting of an additional frequency-dependent
delay term:
\begin{equation}
\Delta_{\rm FDD} = k_{\rm f}\left(f^{\rm SSB}\right)^\zeta,
\end{equation}
\noindent where $k_{\rm f}$ and $\zeta$ define the scale and spectral
index of the term. This term may be used in conjunction with the
stride fit feature to model, for example, delays due to
refraction in the interstellar medium \citep{fc90} or deviations
from the cold plasma dispersion law \citep{pw92}.

\subsubsection{Vacuum propagation delay}
\label{sec:vpd}
The vacuum propagation delay is affected by the distance to the pulsar
and the variation of this with time. The full geometric distance was
derived in Section \ref{sec:geometric}.  The first term from equation
(\ref{eq:geometric}) to be assigned to $\Delta_{\rm IS}$,
$|\vec{R_0}|/c$, is in fact not measurable to even remotely sufficient
precision, but is constant and so can be dropped from the timing
formula. A side effect of this is that epochs of measured pulse
frequencies, times of periastrons, glitches and so on are in fact
retarded by the initial light propagation time, $c|\vec{R_0}|$.  The
remaining parts assigned to the interstellar delay are as follows:
\begin{equation}
s_{\rm IS} = k_\parallel + \frac{|\vec{k_\perp}|^2}{2|\vec{R_0}|}
- \frac{k_\parallel|\vec{k_\perp}|^2}{2|\vec{R_0}|^2}. 
\label{eq:sism}
\end{equation}

The first term on the right hand side of equation (\ref{eq:sism})
simply represents the displacement of the system in the initially radial
direction. This receives potentially significant contributions from
the radial velocity of the pulsar and the radial acceleration in the
gravitational potential of the globular cluster or Galactic
environment \citep{dt91}. The second term describes the so-called
Shklovskii effect \citep{shk70}, whereby initially transverse motion
attains a radial component due to the change in the direction of the
line of sight caused by the transverse motion of the pulsar. That is,
the pulsar moves tangentially while a path of constant distance
describes a circle centred on the SSB. The distance between the two is
the excess propagation length. The third term in corresponds to
secular change in the size of the Shklovskii effect as the radial
motion increases or decreases the distance, and hence changes the
curvature of the line of constant distance just mentioned. An
alternative interpretation of this term is that it is the cumulative
effect upon the Shklovskii term caused by the increase or decrease in the 
apparent proper motion owing to the increasing
transverse component of the initially radial motion (Section
\ref{sec:roemer}). To our knowledge, this effect was first noted
by \citet{van03a}, in the context of its manifestation as an apparent
second derivative of the spin and orbital periods. 


These effects are often neglected in timing formulae, because as
simple constant or secularly increasing Doppler shifts, they are
inseparable from the pulse and orbital periods and their derivatives.
A common procedure is to correct the apparent period derivatives by
estimating the contributions of transverse motion and gravitational
acceleration and subtracting them to obtain the intrinsic value (e.g.\
\citealt{dt91,ctk94}). Alternatively, in certain binary
systems where the true orbital period derivative is
expected to be negligible, the measured derivative can be attributed
entirely to these effects to obtain, in combination with a measurement
of the transverse proper motion, an estimate of $|\vec{R_0}|$
\citep{bb96}. Likewise, a measured second spin frequency derivative
could, in principle, yield a measurement of the radial velocity
\citep{van03a} via the third term of equation (\ref{eq:sism}).

As an alternative to modelling these effects via artificial
contributions to spin and binary parameters, \tempoII\ can optionally
include them directly in the following form:
\begin{equation}
\Delta_{\rm VP} = 
 \frac{v_\parallel}{c}\left(t^{\rm BB}_{\rm a}-t_{\rm pos}\right)
   + \frac{a_\parallel + d_{\rm Shk}|\vec{\mu_\perp}|^2}{2c}
   \left(t^{\rm BB}_{\rm a}-t_{\rm pos}\right)^2
 +  \frac{1}{2c}\left(a_\mu-v_{\parallel{\rm Shkdot}}|\vec{\mu_\perp}|^2
  \right)
     \left(t^{\rm BB}_{\rm a}-t_{\rm pos}\right)^3  ,
\label{eq:vpd}
\end{equation}
\noindent where $v_{\parallel{\rm Shkdot}}$, $a_\parallel$, $d_{\rm
Shk}$ and $a_\mu \equiv \vec{a_\perp}\cdot\vec{\mu_\perp}$ are fittable
parameters that specify the radial velocity, radial acceleration,
Shklovskii distance, and component of acceleration in the direction of
the transverse proper motion.   The terms involving
acceleration arise due to the expansion of $\vec{k}$ in terms of
proper motion, distance and acceleration (equation \ref{eq:k}). The
circular dependence of $\Delta_{\rm VP}$ and $t^{\rm BB}_{\rm a}$ is
resolved by iteration as described in Section \ref{sec:roemer}.

The parameters $a_\parallel$ and $d_{\rm Shk}$ are inseparable, and
highly covariant with a simultaneous change to the spin frequency
derivative and orbital period derivative. It is therefore imperative
that at least one of the first two parameters plus any one of the
remaining three parameters are held fixed at some independently
determined value when fitting. These parameters are included
separately because each of them is potentially subject to external
constraints. The transverse proper motion, $\mu_\perp$, is typically
immune from this covariance due to its appearance in the annual proper
motion term (Section \ref{sec:roemer}). Likewise, $v_{\parallel{\rm
Shkdot}}$ and $a_\mu$ are inseparable, and highly covariant with the
spin frequency second derivative, so only one of these parameters should
be allowed to vary in a fit. They are included separately because
$v_\parallel$ may be determined from $\mu_\parallel$ (Section
\ref{sec:roemer}) and one or more of the distance parameters, while
$\vec{a}$ may be estimated from models of the Galactic rotation and
gravitational potential. The first-order radial velocity term is
discussed in the following section.

\subsubsection{Transformation from SSB to BB coordinate time}
\label{sec:doppler}
The final term contributing to $\Delta_{\rm IS}$ accounts for the
special relativistic time dilation owing to the relative velocities
of the Solar system and binary barycentres:
\begin{equation}
t^{\rm BB}-t_{\rm pos} = \frac{t^{\rm SSB}-t_{\rm pos}}
  {\sqrt{1-|\vec{v}|^2/c^2}},
\end{equation}
where $\vec{v}$ is the relative velocity of the frames. 
This transformation is applied to the arrival time of the pulse at
the BB, so that
\begin{eqnarray}
t^{\rm BB}_{\rm a} &=& t_{\rm a}^{\rm SSB} - \Delta_{\rm ISD} -\Delta_{\rm FDD} -\Delta_{\rm VP} - \Delta_{\rm ES},\;{\rm where} \\
\Delta_{\rm ES} &=& \frac{v^2}{2c^2}\left(t_{\rm a}^{\rm SSB}-t_{\rm pos}- \Delta_{\rm ISD}-\Delta_{\rm FDD} -\Delta_{\rm VP}\right),
\end{eqnarray}
\noindent neglecting terms of order $v^4/c^4$ and higher.
Here the notation $\Delta_{\rm ES}$ refers to the fact that the term
is analogous to the Solar system and binary Einstein delays 
(Sections \ref{sec:einstein} and \ref{sec:binaryeinstein}) and relates
to the relative secular motion of the SSB and BB.

This transformation combines with the much larger $v_\parallel/c$ term
of equation (\ref{eq:vpd}) to effect a Doppler shift
between the two frames: 
\begin{eqnarray}
\Gamma &=& \frac{{\rm d}t^{\rm SSB}_{\rm a}}{{\rm d}t^{\rm BB}_{\rm a}}
 \left(t^{\rm SSB}_{\rm a} = t_{\rm pos}\right)\\
 &=& \frac{1-v_\parallel/c}{\sqrt{1-v^2/c^2}}.
\end{eqnarray}
\noindent An unknown Doppler shift of this kind cannot be
distinguished from a re-scaling of various parameters intended to
refer to the BB frame, meaning that neither
$v_\parallel$ nor $v^2$ can be measured via their contribution to
$\Gamma$. In fact, because the numerical values used for $c$
and $G$ are the same in all frames, the effect of neglecting
$\Delta_{\rm ES}$ is the same as a change of unit system involving
re-scaling of the length mass and time units, and physical tests
involving only ``measured'' (wrong) parameters and these constants
will remain valid \citep{dd86}.

In the timing model presented here (and in contrast to that of
\citealt{dd86}), $\vec{v}$ can in principle be deduced from parameters
measured via other terms of equation (\ref{eq:geometric}). For this
reason it may be included by specifying non-fittable values for the
initial radial velocity $v_\parallel$ and transverse speed $v_\perp$
in the parameter file, which correct for the Doppler effect via equation
(\ref{eq:vpd}) and the following formulation of the relativistic second-order
Doppler shift:
\begin{equation}
\Delta_{\rm ES} = \frac{v_\parallel^2+v_\perp^2}{2c^2}\left(t_{\rm a}^{\rm SSB}-t_{\rm pos}- \Delta_{\rm ISD}-\Delta_{\rm FDD} -\Delta_{\rm VP}\right)
\end{equation} 
\noindent However, it must be remembered that in practice, the
uncertainty of many parameters measured in the frame of the binary
barycentre or pulsar will be dominated by the uncertainties in
$v_\parallel$ and $v_\perp$.

\subsection{The effects of a binary companion}
\label{sec:binary}
The effects of a binary companion are separated into several components:
the geometric Roemer delay, which accounts for excess the vacuum
light travel time due to the Euclidean displacement of the pulsar, a
pseudo-delay that accounts for the aberration of the radio beam by
binary motion, the Einstein delay (combined gravitational redshift
and special relativistic time dilation in the pulsar frame) and
Shapiro delay (gravitational time dilation in the vicinity of
the companion):
\begin{equation}
\Delta_{\rm B} = \Delta_{\rm RB} + \Delta_{\rm AB} + \Delta_{\rm EB}
 + \Delta_{\rm SB}.
\end{equation}

The binary model used by \tempoII\ is based upon those of
\citet{bt76}, Damour \& Deruelle (1986; hereafter DD) \nocite{dd86},
\citet{tw89}, Wex (1998 unpublished contribution to \tempo; see
also \citealt{lcw+01}) and \citet{wex98}, with extra terms as described by
\citet{kop95,kop96}.


\subsubsection{Roemer delay and Kopeikin terms}
\label{sec:binaryroemer}
\label{sec:binarykop}
The variation in the distance of the pulsar due to binary motion contributes
four significant terms to the full geometric path length of equation
(\ref{eq:geometric}):
\begin{equation}
s_{\rm B} = b_\parallel + 
\frac{1}{|\vec{R_0}|}\left(\vec{k_\perp}\cdot\vec{b_\perp}
  - \vec{r_\perp}\cdot\vec{b_\perp} + \frac{|\vec{b_\perp}|^2}{2}\right).
\label{eq:sb}
\end{equation}

The first term of equation (\ref{eq:sb}) is the well-known first-order
geometric delay due to the initially radial component of the displacement 
due to binary
motion. The remaining terms are much smaller in magnitude and have
only proven measurable in a few pulsars to date. Hereafter we shall refer to
these collectively as the ``Kopeikin terms''.

The second term of equation (\ref{eq:sb}) describes changes to the
apparent viewing geometry of the orbital motion, due to the proper
motion of the system \citep{ajrt96,kop96}. An approach taken in the
past to account for this effect is to allow it to be absorbed in
derivatives of the projected semi-major axis, $x$, and longitude of
periastron, $\omega$, with the measured values being converted into
constraints on the position angle of ascending node, $\Omega$, and the
inclination, $i$ (e.g. \citealt{sbm+97,nss01}). Alternatively, the
binary model can be modified to explicitly include linear changes to
$x$ and $\omega$ as parameterised by fittable parameters $\Omega$ and
$i$ (e.g. \citealt{vs04}). Here we take a third approach which we
believe is cleaner: rather than defining the orbital parameters in
a rotating reference frame, we state explicitly that the orbital
parameters refer to the reference frame defined by the ``initial''
geometry at time $t_{\rm BB}$, and hence account for the
effect directly via the second term of equation (\ref{eq:sb}). 

The third term of equation (\ref{eq:sb}) describes the so-called
annual-orbital parallax \citep{kop95}. This may be interpreted either
as the modulation of the Solar system Roemer delay (Section
\ref{sec:roemer}) due to the transverse orbital motion of the pulsar,
or the modulation of the binary Roemer delay (first term of equation
\ref{eq:sb}) due to the transverse orbital motion of the Earth. As
with the secular changes to the apparent viewing geometry, although
the annual-orbital parallax has been modelled in the past by
perturbing the orbital parameters with terms involving
$\vec{r}$ \citep{vbb+01,sns+05}, we choose maintain a
consistent reference frame for the orbital parameters and instead
model the effect directly via the third term of equation (\ref{eq:sb}).

The fourth term of equation (\ref{eq:sb}) is the orbital parallax
\citep{kop95}, which accounts for the fact that transverse binary
motion alters the line of sight and thereby acquires an apparently
radial component: this is the orbital equivalent of the Shklovskii
effect (Section \ref{sec:vpd}). To our knowledge this is yet to be
measured in any pulsar, and naturally cannot be absorbed in
time-varying orbital parameters since the changes themselves occur on
the orbital timescale.

The contributions of the four terms of equation (\ref{eq:sb}) are included
in the timing model as follows:
\begin{eqnarray}
\Delta_{\rm RB} &=& \Delta_{\rm RB\parallel} + \Delta_{\rm KB},\;{\rm where}
\label{eq:binaryroemer}\\
\Delta_{\rm RB\parallel} &=& \frac{b_\parallel}{c}\;\;{\rm and} \\
\Delta_{\rm KB} &=& 
\frac{1}{c}\left(t_{\rm a}^{\rm BB}-t_{\rm pos}\right)\vec{\mu_{\perp \rm B}}\cdot{b_\perp}
- \frac{\vec{r_\perp}\cdot\vec{b_\perp}}{cd_{\rm AOP}}
+ \frac{|\vec{b_\perp}|^2}{2cd_{\rm OP}}.\label{eq:kopeikin}
\end{eqnarray}
\noindent where $d_{\rm AOP}$ and $d_{\rm OP}$ are estimates of the initial
distance which can be either held fixed at some independently
determined value (or omitted from the parameter file, to neglect the
effect), tied to other fittable distance parameters (e.g. $d_{p}$ and/or
$d_{\rm Shk}$), or, given sufficient constraints on the relevant orbital
parameters, fitted to obtain independent distance estimates.
Likewise, $\mu_{\perp \rm B}$ can be either tied to  $\mu_{\perp}$, or
omitted from the parameter file to neglect the effect. Note that these
parameters in fact refer to the quantities as observed in a frame
that is comoving with the BB, in contrast to corresponding
distance and proper motion parameters in preceding sections. However,
the transformation can be neglected for a fractional error of
$O(v^2/c^2) < \sim 10^{-5}$ in those parameters, corresponding to
changes in the time of emission calculation well below our 1-ns goal.

\subsubsection{Post-Newtonian orbital kinematics}
\label{sec:binarykinematics}
Calculation of the various orbital delays depends upon a knowledge of
the displacement $\vec{b}$ of the pulsar from the binary system
barycentre, or at least its radial component $b_\parallel$, depending
on the demanded accuracy. \TempoII\ follows the generalised
post-Newtonian treatment of DD for the calculation of
$\vec{b}$, with the addition of secular derivatives for the
orbital period, eccentricity and projected semi-major axis
after \citet{tw89}. Specifically, from equations (16--17) of DD
\begin{eqnarray}
\vec{b} &=& 
\left(\begin{array}{ccc}\unitvec{e_1}& \unitvec{e_2}& \unitvec{R_0}
\end{array}\right)
\left(\begin{array}{ccc}
 \sin\Omega & -\cos \Omega & 0 \\
 \cos\Omega & \sin\Omega & 0 \\
 0 & 0 & 1
\end{array}\right)
\left(\begin{array}{ccc}
 1 & 0 & 0 \\
 0 & -\cos i & -\sin i\\
 0 & \sin i & -\cos i 
\end{array}\right)
\left(\begin{array}{c}
|\vec{b}| \cos\theta \\ |\vec{b}|\sin\theta \\ 0
\end{array}\right),\;{\rm where} \label{eq:b}\\
|\vec{b}| &=& a(1-e_r\cos u), \\
n\left(t_{\rm e}^{\rm psr}-T_0\right) &=& u - e \sin u, \label{eq:kepler}\\
\theta &=& \omega + A_{e_\theta}(u),\label{eq:binarytheta}\\
A_e(u) &=& 2 \tan^{-1}\left[\left(\frac{1+e}{1-e}\right)^2
                  \tan\frac{u}{2}\right],\\
e_r&=&e(1+\delta_r) \\
e_\theta&=&e(1+\delta_\theta)\\
n&=&\frac{2\pi}{P_{b0}} + \frac{\pi\dot{P_{b}}\left(t_{\rm e}^{\rm psr}-T_0\right)}{P_{b0}^2},\\
\omega &=& \omega_0 + kA_e(u),\label{eq:omega}\\
k&=&\frac{\dot\omega}{n}, \\
e&=&e_0 + \dot{e}\left(t_{\rm e}^{\rm psr}-T_0\right),
\end{eqnarray}
\noindent where $P_{b0}$ and $\dot{P_b}$ are the initial value of the
orbital period and its derivative, $\omega_0$ and $\dot\omega$ are
initial longitude of periastron and the mean of its derivative, $T_0$
is the proper time of periastron, $e_0$ and $\dot e$ are the initial
``proper time eccentricity'' and its derivative, and $\delta_r$ and
$\delta_\theta$ parameterise relativistic deformations of the orbit.
The matrices of equation (\ref{eq:b}) in right-to-left order account for the
inclination of the orbit to the line of sight ($i$), rotation about
the line of sight ($\Omega$), and projection from a
radial--transverse basis to a frame rotationally aligned to the ICRS
(Section \ref{sec:frames}). The matrices are chosen such that for $i=0$, the
angular momentum vector of the orbit is antiparallel to
$\unitvec{R_0}$, and $\Omega$ measures the position angle of the
ascending node with respect to $\unitvec{e_2}$, in the sense of
rotation into $\unitvec{e_1}$.  The definitions of $i$ and $\Omega$
therefore correspond to astronomical convention if $\unitvec{e_1}$ and
$\unitvec{e_2}$ are east and north vectors. This is the case by
default in \tempoII: $\unitvec{e_1}=\unitvec{\alpha}$,
$\unitvec{e_2}=\unitvec{\delta}$.  Alternatively, if ecliptic
coordinates are in use (Section \ref{sec:roemer}),
$\unitvec{e_1}=\unitvec{\lambda}$ and
$\unitvec{e_2}=\unitvec{\beta}$, so that position angles are measured
counter-clockwise from an ecliptic meridian.

Note that the solution to Kepler's equation (\ref{eq:kepler}) involves
the proper time of emission, $t_{\rm e}^{\rm psr}$. This depends upon
the equivalent coordinate time of arrival at the BB, which in turn is
related to the proper time of reception at the observatory by all of
the terms in the timing model, including those involving $t_{\rm
e}^{\rm psr}$. This circular dependence is resolved by starting with
$t_{\rm psr}^{e} = t_{\rm a}^{\rm BB}$, computing the orbital delays,
making an updated estimate of $t_{\rm psr}^{e}$, and iterating this
process until the change in the orbital delay is less than 100 ps.

Using the above expression for $\vec{b}$ and writing it in terms of
its transverse and radial components, the geometric delay of equation
(\ref{eq:binaryroemer}) can be found explicitly. Beginning by
expanding \ $S\equiv|\vec{b}|/a \sin \theta$ and
$C\equiv|\vec{b}|/a\cos\theta$,
\begin{eqnarray}
S &=& \sin\omega \left(\cos u - e_r\right)  + 
           \cos\omega\left(1-e^2_\theta\right)^{1/2}\sin u, \; {\rm and}\\
C &=& \cos\omega \left(\cos u - e_r\right) 
   - \sin\omega\left(1-e^2_\theta\right)^{1/2}\sin u, \label{eq:C}
\end{eqnarray}
\noindent where we have followed the Appendix of DD in making a minor
modification to the definition of $e$ to simplify the expressions. Writing
equation (\ref{eq:b}) in terms of the projections of 
$\vec{b}$ upon $\unitvec{e_1}$, $\unitvec{e_2}$ and $\unitvec{R_0}$ as
a function of $S$ and $C$,
\begin{eqnarray}
b_1 &=& a \left(C \sin\Omega + S\cos\Omega \cos i\right),\\
b_2 &=& a \left(C \cos\Omega - S\sin\Omega \cos i\right),\;{\rm and}\\
b_\parallel &=& a S \sin i .
\end{eqnarray}
The basic Roemer delay may then be written:
\begin{eqnarray}
\Delta_{\rm RB\parallel} &=& x S,\; {\rm where}\label{eq:delta_rbpar}\\
x &=& x_0 + \dot{x}\left(t^{\rm psr}_{\rm e}-T_0\right),
\end{eqnarray}
\noindent and $x \equiv a/c \sin i$ is the projected semi-major axis
as a vacuum light travel time, defined in terms of its initial value,
$x_0$ (DD) and its derivative, $\dot{x}$ \citep{tw89}. For the
Kopeikin terms, following equation \ref{eq:kopeikin},
\begin{eqnarray}
\Delta_{\rm KB} &=& \Delta_{\rm SR} + \Delta_{\rm AOP} + \Delta_{\rm OP},\;{\rm where}\\
\Delta_{\rm SR} &=&   x \left(t_{\rm a}^{\rm BB}-t_{\rm pos}\right)
 \left[(\mu_1\sin\Omega + \mu_2\cos\Omega) C\csc i
    + (\mu_1\cos\Omega - \mu_2\sin\Omega) S\cot i \right],
\label{eq:delta_sr}\\
\Delta_{\rm AOP} &=& -\frac{x}{d_{\rm AOP}} 
\left[(\vec{r}\cdot\unitvec{e_1}\sin\Omega 
            + \vec{r}\cdot\unitvec{e_2}\cos\Omega) C\csc i
    + (\vec{r}\cdot\unitvec{e_1}\cos\Omega 
           - \vec{r}\cdot\unitvec{e_2}\sin\Omega) S\cot i \right],\;{\rm and}
\label{eq:delta_aop}\\
\Delta_{\rm OP} &=& \frac{cx^2}{2d_{\rm OP}}  \left(C^2 \csc^2 i + S^2\cot^2 i\right).
\label{eq:delta_op}
\end{eqnarray}
The proper motion components are defined in terms of fittable
parameters that can optionally float independently to the annual
proper motion parameters (Section \ref{sec:roemer}), in order to allow
the latter to be separated from the secular change of binary viewing
geometry. For equatorial coordinates, $\mu_1=\mu_{\alpha\rm B}$ and
$\mu_2=\mu_{\alpha\rm B}$, while for ecliptic coordinates
$\mu_1=\mu_{\lambda\rm B}$ and $\mu_2=\mu_{\beta\rm B}$. In agreement
with astronomical convention, the parameter $\Omega$ corresponds to
the position angle of periastron in terms of eastward rotation from
north, while the parameter $i$ is defined such that $i=0$ corresponds
to a binary orbital angular momentum vector directed towards the
observer.  As pointed out by \citet{sns+05}, these differ from the
definitions of \citet{kop95,kop96}. We note that some caution is
required in this area owing to variations in definitions in the
literature: for other examples of use of the unconventional definition
of $i$ see e.g. DD; \citet{dt92,wt02,sta04}.


\subsubsection{Aberration}
\label{sec:binaryaberration}
Owing to the relative transverse velocity of the pulsar and the Earth,
the direction of the observer as seen from the pulsar differs from
that which would be measured in a frame co-moving with the observer.
The direction is aberrated according to the Lorentz transformation
that relates the two frames. Under the assumption that the source of
pulses is the periodic rotation of an emission beam, any such change
in the direction of the observer alters the rotational phase
corresponding to radiation received at a given time. Following DD
(equation 27), in \tempoII\ this is converted to an equivalent time
delay:
\begin{equation}
\Delta_{\rm AB} = A\left\{\sin[\omega+A_{\rm e}(u)]+e\sin\omega\right\}
+B\left\{\cos[\omega+A_{\rm e}(u)]+e\cos\omega\right\},
\end{equation}
\noindent where $A$ and $B$ are parameters related to the orientation
of the pulsar spin axis and the size of the orbit (see DD equations 38--39 and DD section 3.2). As discussed by DD, these parameters are highly covariant
with the other binary parameters, and may only be separated on the time scale
of geodetic precession. Note, here the relative motion of the
observatory and the BB is neglected.

\subsubsection{Post-Newtonian delays}
\label{sec:binaryshapiro}
\label{sec:binaryeinstein}
\label{sec:binaryreldel}
In addition to the modified Roemer delay of equation
(\ref{eq:binaryroemer}) and the aberration pseudo-delay, under the
post-Keplerian formalism of DD there are two additional delay terms:
the Einstein delay and the Shapiro delay.

The Einstein delay is the difference between the proper time of
emission and the coordinate time in the quasi-inertial frame of the
binary barycentre:
\begin{equation}
t^{\rm BB} = t^{\rm psr} + \Delta_{\rm EB}.
\end{equation}
\noindent This is due to the combined effect
of gravitational redshift and time
dilation. Following DD and in contrast to the \tempoII\ treatment of
the Solar system Einstein delay, the binary Einstein delay specifically
excludes a linear term by scaling $t^{\rm psr}$ in such a way that the orbital 
period is numerically the same in either time scale. It is expressed
in a theory-independent manner 
in terms of a timing model parameter, $\gamma$; from equation (19) of DD:
\begin{equation}
\Delta_{\rm EB} = \gamma \sin u .
\end{equation}

The Shapiro delay is caused by the curvature of space-time in the
gravitational field of the binary companion. After equation (26) of DD:
\begin{equation}
\Delta_{\rm SB} = -2r\log\left\{1-e\cos u - s
   \left[\sin\omega\left(\cos u - e \right)
         + \left(1-e^2\right)^{1/2} \cos\omega \sin u\right]\right\}.
\end{equation}
Here the fittable parameters are the theory-independent ``range'',
$r$, and ``shape'', $s$. Under general relativity, $s = \sin i$ and
$r=Gm_2/{c^3}$, where $m_2$ is the mass of the binary companion. In
\tempoII, $r$ is expressed in units of $T_\odot = GM_\odot/{c^3}$, so
that its value is numerically equal to the GR companion mass in units
solar masses. ($T_\odot$ is half the light travel time across the
Solar Schwarzschild radius.) An alternative parameterisation is
available, where the parameter $s$ is replaced by the
``\textsc{shapmax}'' parameter, $z_s \equiv -\ln(1-s)$, This derives
from a modification to \tempo\ designed to avoid difficulties in
fitting highly inclined orbits near $|\sin i|=1$ \citep{ksm+06a}. In
this case, the above relation applies after substituting $s =
1-\exp(-z_s)$.

\subsubsection{Assuming general relativity}
\label{sec:binarygr}
The post-Keplerian parameters of Sections \ref{sec:binarykinematics} and
\ref{sec:binaryreldel} ($\dot P_b$, $\dot\omega$, $\delta_r$,
$\delta_\theta$, $\gamma$, $s$, $m_2$) can either be
specified directly, or they can be derived from a reduced set of
physical parameters based on the assumption that the predictions
made by general relativity are correct.  These parameters are $m_2$ and
$M\equiv m_1+m_2$, the total system mass (in addition to the Keplerian
parameters, $P_{b0}$, $\omega_0$, $e$, $T_0$, and $x_0$). After
equations 8.48--8.55 of Lorimer \& Kramer (2005)\nocite{lk05},
\begin{eqnarray}
\dot\omega^{\rm GR} &=& 3T_\odot^{2/3} n^{5/3} \frac{M^{2/3}}{1-e^2}, \\
\gamma^{\rm GR} &=& T_\odot^{2/3} n^{-1/3} e \frac{m_2\left(m_1+2m_2\right)}{M^{4/3}},\\
r^{\rm GR} &=& T_\odot m_2, \\
s^{\rm GR} &=& \sin i = T_\odot^{-1/3} n^{2/3} x \frac{M^{2/3}}{m_2},\\
\delta_r^{\rm GR} &=& T_\odot^{2/3} n^{2/3} \frac{3m_1^2+6m_1m_2+2m_2^2}{M^{4/3}},\\
\delta_\theta^{\rm GR} &=& T_\odot^{2/3} n^{2/3}\frac{\frac{7}{2}m_1^2+6m_1m_2+2m_2^2}{M^{4/3}},\; {\rm and} \\
\dot P_{b}^{\rm GR} &=& -\frac{192\pi}{5} T_\odot^{5/3} n^{5/3}\frac{m_1 m_2}{M^{1/3}}\frac{1+73e^2/24 + 37e^4/96}{(1-e^2)^{7/2}}.
\end{eqnarray}



Although
these equations account fully for the general relativistic rates of
change of orbital period and longitude of periastron, extra variation
in these parameters is allowed for via the parameters $\dot{P}_{bx}$ and
$\dot\omega_x$:
\begin{eqnarray}
n^{\rm GR} &=& \frac{2\pi}{P_{b0}} + 
 \frac{\pi\left(\dot{P}_b^{\rm GR}+\dot{P}_{bx}\right)\left(t_{\rm e}^{\rm psr}-T_0\right)}{P_{b0}^2}\; {\rm and} \\
\omega^{\rm GR} &=& \omega_0 
          + \left(k^{\rm GR}+\frac{\dot\omega_x}{n}\right)A_e(u).
\end{eqnarray}
\noindent Note that $\dot\omega_x$ is not intended to model the secular
change in orbital viewing geometry due to proper motion: this is modeled
by a separate term in the timing model, causing $\omega$ to be defined
in a consistent, non-rotating reference frame 
(Section \ref{sec:binaryroemer}).

Under the condensed parameterisation of this section, the orientation of
the spin axis may be separated from the orientation of the orbit
normal vector to yield a physical parameterisation of the aberration
delay (DD).  Following equations 8.56--8.57 of Lorimer \& Kramer (2005):
\begin{eqnarray}
A^{\rm GR} &=& \frac{T_\odot^{-1/3}}{(2\pi)^{2/3}}
  \frac{m_2}{\nu P_b^{1/3}(1-e^2)^{1/2}M^{2/3}} \sin\chi \csc\xi,\\
B^{\rm GR} &=& -\frac{T_\odot^{-1/3}}{(2\pi)^{2/3}}
  \frac{m_2}{\nu P_b^{1/3}(1-e^2)^{1/2}M^{2/3}} \cos i \cos\chi \csc\xi
\label{eq:BGR},
\end{eqnarray}
\noindent where $\nu$ is the pulsar spin frequency (Section
\ref{sec:phase}), $\xi$ is the inclination of the pulsar spin axis,
and $\chi$ is its position angle\footnote{These are related to the
$\eta$ and $\lambda$ of \citet{dt92} and Lorimer \& Kramer (2005) via
$\chi=-\eta$, $\xi=\pi-\lambda$.  The difference is to ensure the use
of the conventional definition of position angle and inclination, see
Section \ref{sec:binarykinematics}.}, relative to the position angle
of ascending node, $\Omega$. Here $\chi$ and $\csc \xi$ are the
fittable parameters. Note that because $\cos i$ is only ambigiously
constrained by the Shapiro delay term ($\cos i = \pm\sqrt{1-\sin^2
i}$), $\chi$ too is subject to ambiguity.  To evaluate equation
(\ref{eq:BGR}), \tempoII\ assumes that $\cos i \geq 0$.  Values of
$\chi$ determined on this basis are therefore indistinguishable under
the transformation $\chi \rightarrow \pi - \chi$. Only if $i$ is
measurable via the Kopeikin terms can this degeneracy be broken.  A
second, unbreakable degeneracy exists due to the fact that aberration is only
sensitive to the projection of the transverse orbital velocity
upon the spin axis: $\xi \rightarrow \pi-\xi$.

\subsubsection{Nearly circular orbits}
\label{sec:ell1}
For low-eccentricity orbits, $\omega$ and $T_0$ are highly covariant,
which complicates the analysis of the uncertainty in these parameters.
This problem can be avoided by parameterising the motion in terms
of the Laplace-Lagrange parameters,
\begin{eqnarray}
\eta &\equiv& e\sin\omega\;\; {\rm and} \\
\kappa &\equiv& e \cos\omega,
\end{eqnarray}
\noindent and the time of ascending node,
\begin{equation}
T_{\rm asc} \equiv T_0 - \omega/n
\end{equation}
\noindent (Wex 1998 unpublished contribution to \tempo\ as ``ELL1'' binary
model). Following \citet{lcw+01},  the following approximation
to equation (\ref{eq:delta_rbpar}) applies, to order $e$:
\begin{eqnarray}
\Delta_{RB\parallel}^{\rm LL} &=& x^{\rm LL}S^{\rm LL},\;{\rm where}\\
S^{\rm LL} &=& \sin\Phi + \frac{\kappa}{2}\sin 2\Phi 
               - \frac{\eta}{2}\cos 2\Phi \label{eq:SLL},\\
\Phi &=& n\left(t^{\rm psr}_{\rm e}-T_{\rm asc}\right), \\
x^{\rm LL} &=& x_0 + \dot{x}\left(t^{\rm psr}_{\rm e}-T_{\rm asc}\right), \\
\eta &=& \eta_0 + \dot{\eta}\left(t^{\rm psr}_{\rm e}-T_{\rm asc}\right),
\;{\rm and}\\
\kappa &=& \kappa_0 + \dot{\kappa}\left(t^{\rm psr}_{\rm e}-T_{\rm asc}\right).
\end{eqnarray}
\noindent Here $\eta_0$, $\dot{\eta}$, $\kappa_0$, $\dot{\kappa}$ and
$T_{\rm asc}$ replace $e_0$, $\dot{e}$, $\omega_0$, $\dot{\omega}$ and $T_0$
as fittable parameters, while $\delta_\theta$ and $\delta_r$ are neglected.

The original ELL1 model as presented by \citet{lcw+01} did
not include the Kopeikin terms (Section \ref{sec:binaryroemer}).
In order to include them, an expression for $C$ (equation \ref{eq:C}) is
required. To the same accuracy as equation (\ref{eq:SLL}), we
find
\begin{equation}
C^{\rm LL} =  \cos\Phi + \frac{\kappa}{2}\cos 2\Phi 
               + \frac{\eta}{2}\sin 2\Phi,
\end{equation}
\noindent neglecting a constant offset of $3a\kappa/2$ (analogous to an
offset of $3x\eta/2$ dropped by \citealt{lcw+01}). Substitution
into equations (\ref{eq:delta_sr}), (\ref{eq:delta_aop}) and (\ref{eq:delta_op})
yields the Kopeikin terms under this parameterisation.

Under the Laplace-Lagrange parameterisation, the Shapiro delay is
computed as follows \citep{lcw+01}:
\begin{equation}
\Delta_{\rm SB}^{\rm LL} = -\frac{2Gm_2}{c^3}\ln\left(1-\sin i \sin\Phi\right),
\end{equation}
\noindent where either of the parameterisations ($s$, $z_s$) of $\sin i$
of Section \ref{sec:binaryshapiro} may be used.

The aberration delay becomes (Wex 1998 unpublished):
\begin{equation}
\Delta_{\rm AB}^{\rm LL} = A\sin\Phi + B\cos\Phi.
\end{equation}

The Einstein delay is neglected if this parameterisation is chosen.

\subsubsection{Main sequence companions}
\label{sec:binaryms}
The orbits of pulsars with main sequence binary companions show
deviations from a Keplerian orbit attributed to spin-orbit coupling
\citep{lbk95}. \tempo\ includes two binary models for taking these
into account: the BTJ model, which induces steps (``jumps'') in several
parameters at specified epochs (typically, at each periastron), and the
MSS model, which includes orbital and secular variations to the inclination
angle \citep{wex98}. The insertion
of jumps is straightforward and need not be discussed further. The
MSS model is included in the \tempoII\ model by changing the manner in
which $\dot x$ alters $x$, and including second period derivatives in
$x$ and $\omega$:
\begin{eqnarray}
x^{\rm MSS} &=& x_0 + \frac{\dot x A_e(u)}{n} 
   + \frac{\ddot x\left(t^{\rm psr}-T_0\right)^2}{2},\;{\rm and} \\
\omega^{\rm MSS} &=& \omega_0 + kA_e(u) + 
  \frac{\ddot\omega\left(t^{\rm psr}-T_0\right)^2}{2}.
\end{eqnarray}

\subsubsection{Emulating other models}
\label{sec:oldbinary}
As noted in Section \ref{sec:binarykinematics}, the orbital kinematics represent
an inverse problem, because the motion is parameterised by the pulsar
proper time, which itself depends on the orbital propagation delay.
The standard approach in \tempoII\ is to iterate until convergence.
However, in order to reproduce results from the implementations of the
\citet{bt76} or DD models, a facility is provided to alter this
behaviour. Both of these models begin with the zeroth-order
approximation that $t^{\rm psr}_e=t^{\rm BB}_a$ to compute the various
delay terms, then add correction terms to the zeroth-order
approximation to the combined Roemer and Einstein delay,
$\Delta_{\rm RE}^0$, which is expressed
(after DD equations 46--48) as follows:
\begin{eqnarray}
\Delta_{\rm RE}^0 &=& \alpha_{\rm B}(\cos U - e_r) + (\beta + \gamma)\sin U,\; {\rm where}\\
\alpha_{\rm B} &=& \frac{a \sin i}{c} \sin \omega,\\
\beta &=& \frac{a \sin i}{c}\left(1-e_\theta^2\right)^{1/2}\cos\omega, {\rm and}\\
U - e_{\rm T}\sin U &=& n(t^{\rm BB}-T_0).
\end{eqnarray}
\noindent Note that the Kopeikin terms (equation \ref{eq:kopeikin}) are neglected.

The expansion provided by DD is
\begin{eqnarray}
\Delta_{\rm RE}^{DD} &=&  \Delta_{\rm RE}^0\left[
1 - \hat{n}\Delta_{\rm RE}^{0'} 
+ \hat{n}^2\left(\Delta_{\rm RE}^{0'}\right)^2
+ \frac{1}{2}\hat{n}^2\Delta_{\rm RE}^{0}\Delta_{\rm RE}^{0'''}
- \frac{1}{2}\frac{e\sin U}{1-e\cos U}\hat{n}^2
    \Delta_{\rm RE}^{0}\Delta_{\rm RE}^{0''}\right], {\rm where}\\
\hat{n} &=& \frac{n}{1-e\cos U},\\
\Delta_{\rm RE}^{0'} &=& -\alpha_{\rm B}\sin U + (\beta + \gamma)\cos U, {\rm and}\\
\Delta_{\rm RE}^{0''}&=& -\alpha_{\rm B}\cos U - (\beta + \gamma)\sin U.
\end{eqnarray}

The approximation provided by \citet{bt76} is equivalent to:
\begin{eqnarray}
\Delta_{\rm RE}^{\rm BT} &=&\Delta_{\rm RE}^0\left(1-\hat{n}\Delta_{\rm RE}^{0'}\right).
\end{eqnarray}
\noindent 
\noindent In addition, \citet{bt76} neglected $\Delta_{\rm BS}$, $\Delta_{\rm BA}$, 
$\delta_r$ and $\delta_\theta$, and the orbital modulation of $\omega$:
that is, in contrast to equation (\ref{eq:omega}),
\begin{equation} 
\theta^{\rm BT} = \omega_0+\dot\omega\left(t^{\rm psr}-T_0\right).
\end{equation}
\noindent Keeping the DD formulation of $\omega$ but the BT formulation
of the other terms allows for the emulation of the BT+ model of \tempo\ 
\citep{dt92}.

Additionally, it is possible to emulate the \tempo\ implementation of
the Laplace-Lagrange parameterisation (Section \ref{sec:ell1}), which 
made an approximation to the Roemer delay analogous to the DD treatment:
\begin{eqnarray}
\Delta_{\rm RB}^{\rm ELL1} &=& \Delta_{\rm RB}^{\rm ELL1-0} 
\left[1-n\Delta_{\rm RB}^{\rm ELL1-0'}
+\left(n \Delta_{\rm RB}^{\rm ELL1-0'}\right)^2
+\frac{1}{2}n^2\Delta_{\rm RB}^{\rm ELL1-0}\Delta_{\rm RB}^{\rm ELL1-0''}
\right],\;{\rm where}\\
\Delta_{\rm RB}^{\rm ELL1-0} &=& \frac{x}{c}
\left(\sin\Phi^0 + \frac{\kappa}{2}\sin 2\Phi^0
               - \frac{\eta}{2}\cos 2\Phi\right), \\
\Delta_{\rm RB}^{\rm ELL1-0'} &=& x \cos \Phi, \\
\Delta_{\rm RB}^{\rm ELL1-0''} &=&-c \sin\Phi,\; {\rm and}\\
\Phi^0 &=& n(t^{\rm BB}_{\rm e}-T_{\rm asc}).
\end{eqnarray}

\subsection{Pulse phase model}
\label{sec:phase}
The final component of the timing model is the evolution of the
phase of the pulse sequence, relative to the proper time $t^{\rm psr}$
of the pulsar centre of mass (equation \ref{eq:top}). In most cases,
a simple Taylor series expansion is used:
\begin{equation}
\phi(t) = \sum_{n\geq1} \frac{\nu^{(n-1)}}{n!} 
       \left(t^{\rm psr}_{\rm e}-t_{\rm P}\right)^n + \phi_0.
\label{eq:phaseev}
\end{equation}
\noindent The frequency derivative terms ($\nu^{(n)}$) are the
fittable parameters, while the epoch, $t_{\rm P}$, at which
$\dot{\phi}=\nu$ is set by the user. Absolute phase alignment is
achieved by means of $\phi_0$, which is not specified
directly. Instead, in order to protect against changes in $t_{\rm P}$
or the choice of barycentric time scale, $\phi_0$ is defined in terms
of a reference time of arrival for a specified observing site and
observing frequency.  This time of arrival is chosen such that it is
close to the centre of the time span of the fitted data, and has a
predicted pulse phase that is exactly integer.

Note that under the assumption that the basic pulse train is a result
of a rotating emission beam, the time of zero phase is in fact
dependent on the position of the observer. Of all the sources of
relative motion (Earth orbital, binary orbital and secular motion and
acceleration), only the secular motion of the pulsar (or binary
barycentre) is significant given our accuracy goal of 1~ns. However,
since the transverse speed is nearly constant, so is the rate of
change of apparent pulse phase, meaning that the effect cannot be
separated from the basic pulse frequency, $\nu$. For this reason, the
effect is neglected in \tempoII. Likewise, the emitted beam is subject
to aberration due to the relative velocity of the pulsar, however the
constant component of this velocity contributes only a constant phase
shift that is absorbed in $\phi_0$. On the other hand, the binary
motion induces a variable aberration, which for simplicity is included
in $\Delta_B$ as a pseudo-propagation delay $\Delta_A$ (Section
\ref{sec:binaryroemer}).

Though the pulse frequency of most pulsars shows a continual gradual
slow-down, attributed to magnetic braking of a spinning neutron star,
certain pulsars exhibit anomalies known as ``glitches''. These
events are generally well-modelled by permanent increments to the phase,
frequency and frequency first derivative, in addition to a
frequency increment that later decays exponentially to zero:
\begin{equation}
  \phi_{\rm g}= \left\{
\begin{array}{ll}
\begin{array}{l}
\Delta \phi + \Delta \nu (t^{\rm psr}_{\rm e}-t_{\rm g})  + 
\frac{1}{2}\Delta \dot{\nu} (t^{\rm psr}_{\rm e}-t_{\rm g})^2 + \left(1 - e^{-(t^{\rm psr}_{\rm e}-t_{\rm g})/\tau}\right) \Delta \nu_t (t^{\rm psr}_{\rm e}-t_{\rm g}) 
\end{array}
  & t^{\rm psr}_{\rm e} \geq t_{\rm g} \\
0 & t^{\rm psr}_{\rm e} < t_{\rm g} .
\end{array} \right.
\end{equation}
An arbitrary number of glitches may be modelled in this manner, the
corresponding values of $\phi_{\rm g}$ being added to the basic
Taylor series of equation (\ref{eq:phaseev}) to give the predicted evolution
of pulse phase with time.

\section{Accuracy Estimates}
\label{sec:accuracy}
\subsection{Geometric propagation delay}
Errors are introduced to the geometric part of the propagation delay
in two ways: through the neglect of various terms of equation
(\ref{eq:geometric}), and through approximations made in the
computation of other terms. Cases of the latter are discussed
below. The neglected terms are the $O(|\vec{R_0}|^{-3})$ terms, and
all those involving the expansion of the third, second, and fourth
terms in the second set of parentheses, except for the secular and
annual changes to the Shklovskii effect and the secular change to the
annual proper motion. Based upon measured parameters and a timing
campaign spanning 20 years, we have estimated the likely magnitude of
the neglected terms for all pulsars in the current ATNF pulsar
catalogue having spin periods shorter than 30 ms (slower pulsars being
unlikely to yield high precision timing measurements). The largest of
the neglected terms are the change in annual proper motion due to
transverse acceleration ($\sim$ 1 ns for PSR B1620$-$26), second-order
secular changes to the apparent binary orbital viewing geometry ($\sim
600$ ps for PSR J0437$-$4715), the $O(|\vec{R_0}|^{-3})$ terms
involving secular motion ($\sim 600$ ps for PSR B1257+12), the binary
modulation of the Shklovskii effect ($\sim 300$~ps for PSR
J0437$-$4715) and the second-order annual proper motion ($\sim 100$~ps
for PSR J0437$-$4715). 

It is possible that pulsars yet to be discovered may yield larger
errors than the examples listed here, if they are in wider binary
orbits, are closer to the Earth, have larger proper motions, and/or
experience greater secular acceleration. The current sample is not
likely to be biased in relation to proper motion, and any bias in
relation to distance will tend to select closer pulsars due to the
dependence of detection upon flux density.  Highly accelerated pulsars
are difficult to detect, however the level of acceleration at which
this becomes problematic is many orders of magnitude larger than may
be experienced a the Galactic or intracluster environment. Binary
acceleration also hampers detection, but for the known population the
lack of any strong dependence of companion mass upon orbital
separation means that this selection effect biases the sample towards
large orbital separations rather than small. However, it is not
inconceivable that the present sample has missed an entire class of
millisecond pulsars that experience strong binary acceleration due to
very massive companions in wide orbits.  The discovery of such pulsars
may necessitate the re-inclusion of some of the dropped terms of the
geometric propagation delay.
 
\subsection{Clock Corrections}
\label{sec:clkcorr}
\begin{figure}
   \begin{center}
    \includegraphics[width=0.7\hsize]{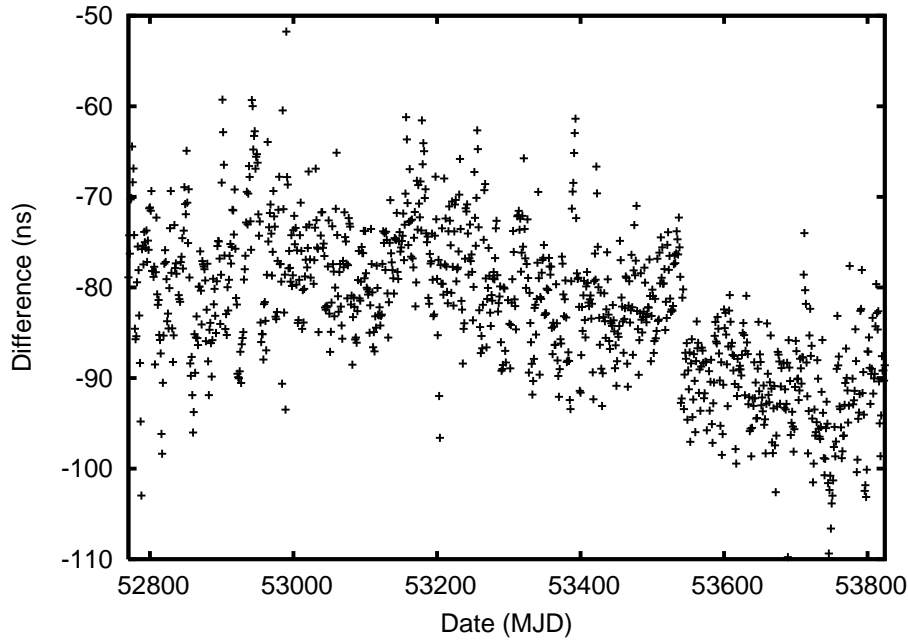}
   \end{center}
  \caption{Round-trip time transfer error involving independent clock
monitoring systems at Parkes observatory. The terms in each correction
are UTC(GPS)$-$UTC(PKS) (``Totally Accurate Clock'' system),
UTC$-$UTC(GPS) (Circular T), UTC$-$UTC(AUS) (Circular T) and
UTC(AUS)$-$UTC(PKS) (GPS common view system). }
\label{fig:gps_cv}
 \end{figure}

Clock corrections applied by \tempoII\ may suffer from time-transfer
errors present in the clock offset files provided by the user, owing
either to incorrect tabulated values, or to variations in the clock
offset on time scales shorter than the tabulation interval, over which
linear interpolation is performed. In most instances, the dominant
source of these errors will be the first step of correction, involving
the local observatory clock, since observatory time standards and time
transfer systems are typically less accurate than those involved in
GPS and UTC time keeping. As an example of the level at which time
transfer errors may occur, we note that differencing of two independent
time transfer systems at the Parkes observatory leads to an error term
with a root-mean-squared (rms) deviation of 8 ns over the last three
years (Fig. \ref{fig:gps_cv}).  Time transfer to UTC from GPS or one
of the national time standards contributing to UTC also leads
significant error, with a typical rms of $\sim 5$ ns \citep{lmpt05}.

\begin{figure}
   \begin{center}
    \includegraphics[width=0.7\hsize]{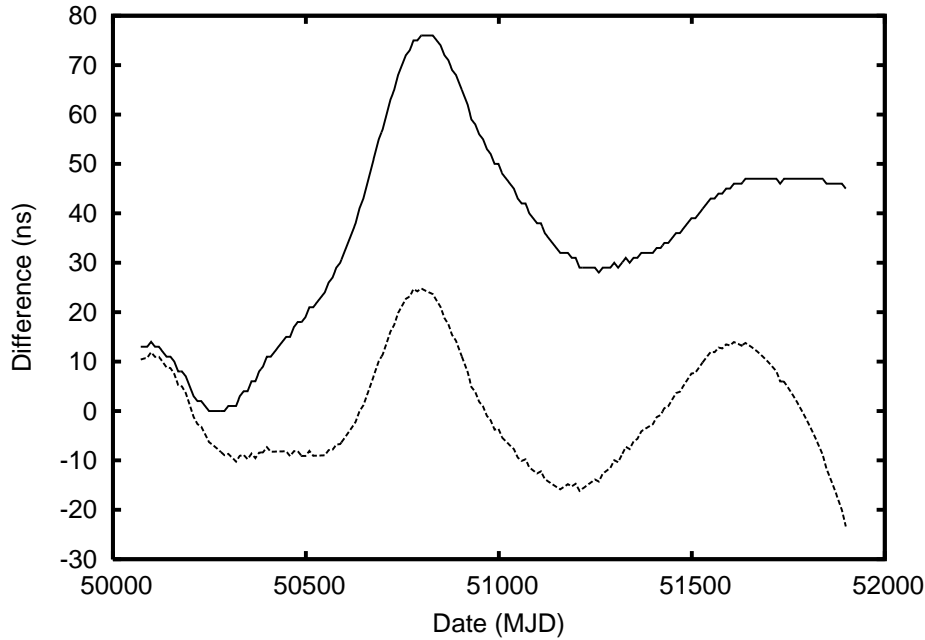}
   \end{center}
  \caption{Difference between two realisations of TT, TT(BIPM2005) and
TT(BIPM2001) before (solid line) and after (dashed line) removing a fitted
function including a quadratic and an annual sinusoid, to account for
the presence of fitted functions of this form in the timing model. The
plot spans the final five years covered by TT(BIPM2001).}
\label{fig:tt01_05}
 \end{figure}

In addition to clock correction errors, the ``corrected'' time of
arrival, as measured against a specified realisation of TT will 
differ from that measured against an ideal TT, owing to instability
in the contributing atomic clock systems. An indication of the
level at which instability contributes can be obtained by means of
the difference between the latest revision of the BIPM realisation
of TT, TT(BIPM2005), with the 2001 version of the same timescale.
As shown in Fig. \ref{fig:tt01_05}, differences exist at the $\sim 10$~ns
level, even after allowing for the absorption of a fraction of the
variation by other terms in the timing model. 

\subsection{Solar system Einstein delay}
The time dilation integral of \citet{if99} (equation
\ref{eq:timedilation}) includes only terms of order $1/c^2$, plus a
mean rate correction for higher-order terms. Inaccuracies in the Solar
system ephemeris lead to an estimated uncertainty of 100~ps in the
time integral, and $10^{-16}$ in its derivative. The remaining
higher-order terms have a maximum amplitude of 33~ps
\citep{fuk95}. The effect upon the spatial coordinates of the
4-dimensional transformation relating the GCRS and BCRS is neglected.
The neglected length contraction is of the order $10^{-8} \vec{s}/c$,
corresponding to up to 200 ps of excess propagation time.

\subsection{Atmospheric delays}
Equation (\ref{eq:hydrostatic_zenith_delay}) predicts the hydrostatic zenith
delay to an accuracy of 0.5 mm under conditions of hydrostatic
equilibrium, with variations of up to 20 mm possible under extreme
weather conditions \citep{dhs+85}. These correspond to timing errors
of $1.7$ -- $67$ ps $\cdot \csc \Theta$. In addition, the accuracy
Niell mapping function degrades at small elevation angles
\citep{nie96}, leading to $\sim 30$~ps of error at $\Theta =
5^\circ$. Should the user not provide tabulated surface atmospheric
pressure, the use of a constant value leads to errors of the order of
1 ns $\cdot \csc \Theta$. Likewise, lack of tabulated values for the
zenith wet delay results in error at the level of $\sim 1$ ns$\cdot
\csc \Theta$.

As noted in Section \ref{sec:atm}, the dispersive component of the
ionosphere causes propagation delays of the order of \\
7$-$130~ns~$\cdot(f/1\;\mathrm{GHz})^{-2}$, where $f$ is the
observing frequency. Such effects must be removed by fitting for a 
time-variable dispersion measure with the aid of multi-frequency
observations.

\subsection{Solar system Roemer delay}
\label{sec:accuracy:roemer}
The Roemer delay depends on the position of the observatory in the
terrestrial frame, provided as input to \tempoII. An accuracy of 30 cm
is needed to meet the 1~ns criterion. In many cases the
position is known to an accuracy of a few centimetres, by means of
VLBI or other geodesy techniques.  In addition, most sites are subject
to continental drift at a rate of a few centimetres per year, which is
neglected by \tempoII.  This may be remedied in future when the
sensitivity and time span of observational data dictates it.

The transformation of the observatory position from the terrestrial to
celestial reference frame depends on a knowledge of the Earth
orientation parameters. Our requirement of 1~ns precision corresponds
to maximum $10$ mas of angular displacement on the surface of the
Earth, or 600~$\mu$s in UT1, the time scale tied to Earth
rotation\footnote{Errors in UT1 relate in fact to a position offset,
and should not be confused with errors that appear directly in the
timing model (such as clock corrections). This is why the precision
demanded of UT1 is some 600,000 times higher than the required timing
accuracy.}.  \tempoII\ uses the ``C04'' Earth orientation parameter
series of the IERS, which exceeds the required precision for all dates
after 1980.  The effects of diurnal and subdiurnal tides are removed
from the series, resulting in an additional error of at most $0.7$ mas
in the location of the pole, and $^{+50}_{-80}\; \mu$s in UT1. The
IAU2000B precession-nutation model used by \tempoII\ is accurate to 1
mas.

As noted in Section \ref{sec:roemer}, \tempoII\ applies transformations
to the input and output values of the Solar system ephemeris that are
specific to recent JPL ephemerides. Should a different ephemeris be
specified, it is important that the offset of the input time
parameter, and the scale of input time and output spatial vectors
match those of DE405, as specified by \citet{if99}. Likewise, it is
important that the ephemeris coordinate frame is rotationally aligned
with the ICRS, for while the gross effect of a frame rotation on
$\vec{r_\oplus} \dotprod \unitvec{R_0}$ could be absorbed in the fitted
pulsar position, the resultant changes in $\unitvec{R_0}$ will induce an
error in $\vec{s}\dotprod \unitvec{R_0}$ with a diurnal signature. For
example, the coordinate frame of the earlier DE200 ephemeris is offset
from the ICRS by $\sim$14 mas \citep{fcf+94}, resulting in up to $\pm 1.4$~ns
of error.

Ultimately the accuracy of the Roemer delay hinges upon the chosen
Solar system ephemeris. Preliminary investigations indicate that
errors in the Solar system ephemeris may be significant and indeed
measurable in pulsar timing data.

\subsection{Solar system Shapiro delays}
\tempoII\ includes the first-order Shapiro delay due all bodies for
which the excess delay exceeds 0.5 ns (see Paper I).
The next largest effect is due to Mars, with 60 ps of delay. The
second order term is included only for the Sun. The next largest
effect is due to Jupiter, with a maximum amplitude of 5~ps. Further
terms appear for the Sun at the 500 ps level \citep{rm83}.

\subsection{Interstellar propagation delays}
The treatment of \tempoII\ is precise in the limit of cold plasma
dispersion with negligible refraction. In practise, turbulence in the
interstellar medium leads to stochastic refraction and scattering of
incident radio waves, giving rise not only to familiar scintillation
effects, but also to frequency-dependent variable propagation
delays \citep{fc90} . Such delays arise both due to differences in the
propagation path length, and also due to induced errors in the
computed Roemer delay, owing to the variations in apparent source
direction. According to simulations conducted by \citet{fc90}, the
amplitude of these terms in the selected pulsars is of the order of
hundreds of nanoseconds, for observations conducted at $\lambda \sim
20$~cm.  
For future, more sensitive telescopes, these
errors will almost certainly dominate over radiometer noise unless
steps are taken to correct for the effect or avoid it by observing
at higher frequencies.

\subsection{Transformation from SSB to BB frame}
The transformation from the relativistic coordinate frames of the SSB
and BB is modelled as a rotationless Lorentz transformation. This is a
pure special relativistic transformation that assumes that the
space-time metric tensors are completely determined by bodies in the
Solar system and binary system respectively. Under general relativity
this assumption will be violated by the presence of gravitational
waves from external sources, causing unmodelled effects in the pulse
times of arrival. Indeed, the detection of the gravitational wave
background is one of the main aims of contemporary pulsar timing
campaigns.  \citet{jhlm05} conducted simulations of a number of
different hypothetical datasets to test the detectability of a
gravitational wave background at the level predicted by theory. A
significant detection was predicted for several simulated datasets of
a standard optimistically within reach of current observing programs,
for which the minimum detectable systematic effect (as indicated by
the rms measurement error divided by the square root of the number
of independent observations) is one to several nanoseconds.

\subsection{Binary orbital effects}
The principal source of error in the formulation of the timing delay
due to the presence of a binary companion is the use of equations of
motion derived at only the first post-Newtonian order \citep{dd86}. To
date no complete timing formula accurate at the second post-Newtonian
(2PN) order has been constructed, although the orbital motion has been
derived by \citet{wex95}, including a proper-time representation that
thereby provides the 2PN Einstein delay. The largest discrepancies
are due to the 2PN contribution to the rate of periastron
advance, and the effect of spin-orbit interaction on the periastron
advance and orbital orientation \citep{dt92}, though these effects can
be absorbed in changes to $\dot\omega$ and $\dot x$.  The 2PN
corrections to the quasi-periodic Einstein delay are of the order 80
ns for PSR B1913+16 \citep{wex95}. Additional unmodelled effects
include the modification to the Shapiro delay due to the
gravitomagnetic field of a spinning companion \citep{lw97}, and the
effect of gravitational light bending on the apparent rotational phase
of the pulsar beam \citep{sch90,dk95}, which cannot be separately
measured \citep{wk99}. For double-neutron-star binaries, all of these
effects are quite large compared to our 1 ns accuracy goal, especially
for nearly edge-on orbits, however it so happens that the binary
evolutionary history of such systems rarely results in a pulsar that
is spinning rapidly enough to provide timing measurements of high
accuracy. 

Only since the discovery of PSR J0737$-$3039A, a 22-ms pulsar in a
highly inclined compact orbit with a second pulsar \citep{lbk+04}, has
the measurability of these effects been seriously considered. It is
expected that the 2PN contribution to $\dot\omega$ will be measurable
in coming years \citep{ksm+06}. Although this effect is easily
accommodated in the existing binary model, we understand that efforts
are underway to develop a fully consistent 2PN timing formula.
Incorporation of this formula in \tempoII\ would be highly
desirable for assurance that the model will continue to be accurate
at the level demanded by ever-decreasing measurement errors and
potentially more extreme binaries discovered in coming decades.

It should be noted that owing to approximations made in the orbital
kinematics, the Laplace-Lagrange parameterisation is only valid
for orbits of low eccentricity. The error term is of order
$xe^2$ \citep{lcw+01}.

\section{Differences from Tempo}
\label{sec:differences}
\subsection{Tropospheric delays}
\tempo\ makes no attempt to model non-dispersive atmospheric delays.
These are dominated by the hydrostatic delay (Section \ref{sec:atm}),
which is a mainly diurnal term of up to tens of nanoseconds in amplitude,
depending on the elevation angle distribution.

\subsection{Einstein delay}
\label{sec:tempo_einstein}
\begin{figure}
   \begin{center}
    \includegraphics[width=0.7\linewidth]{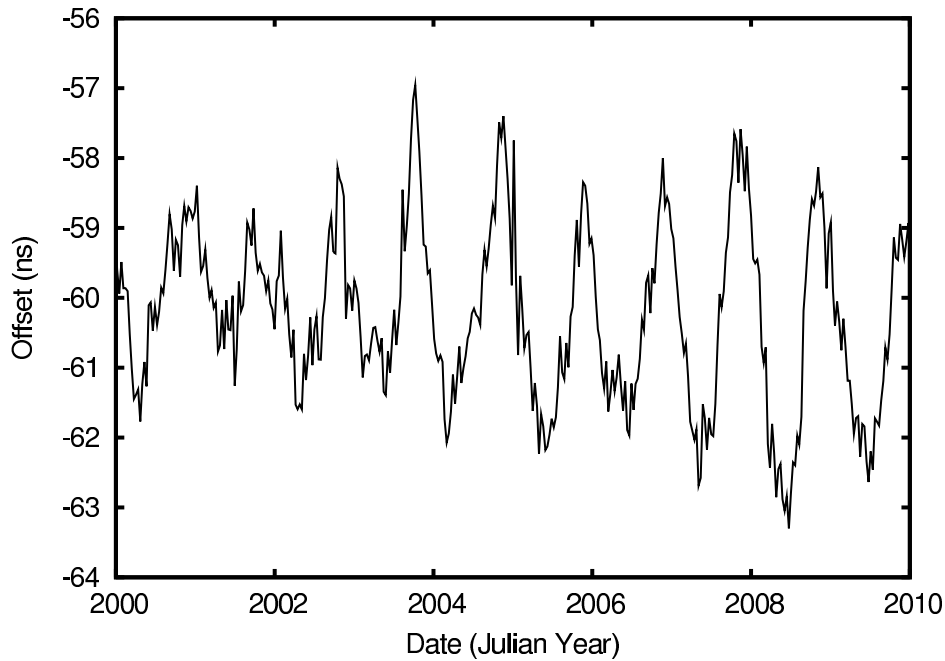}
   \end{center}
  \caption{Difference between timing models of \tempoII\ and \tempo\ owing
to the use of different numerical approximations to the Solar system
time dilation integral.}\label{fig:tephdiff}
 \end{figure}

To compute a version of the time dilation integral of equation
(\ref{eq:timedilation}), \tempo\ uses the series approximation of
\citet{fb90c}, since shown to be significantly in error
\citep{fuk95,if99}.  An example of the discrepancy is shown in
Fig. \ref{fig:tephdiff}, which plots the difference in barycentric
time as computed by \tempoII\ and \tempo. The comparison is
made in terms of the now-obselete Barycentric Dynamical Time (TDB, see
below), as used by \tempo. The difference consists mainly of a
constant offset of $\sim 60$ ns, an annual term of $\sim 2$ ns
amplitude, and a small noise-like contribution due to truncation of
the internal Chebyshev representation used by \tempo. The constant
offset is immaterial since the zero point of TDB has never been
well-defined. It arises for the most part due to the use by
\citet{if99} of an erroneous estimate of the offset between TDB and
$T_{\rm eph}$, with an additional $\sim 6$~ns due to software errors
in early versions of the \citet{fb90c} Fortran code, which only
manifest on certain computer systems (P. Wallace, personal
communication). The annual term, probably due to inaccuracies in the
\citet{fb90c} series, will act to corrupt the source position at
the 0.5 microarcsec level, as fitted by \tempo.

It should also be noted that the time integral actually computed by
\citet{fb90c} and \citet{if99} includes a negative term in the
integrand designed to give a long-term zero mean value of the
integral.  Unlike \tempoII, \tempo\ does not add a counteracting linear
term to the result, causing an overall difference in the scale of the
barycentric coordinate timescale. In the case of \tempo\ the resultant
time refers to the $T_{\rm eph}$ timescale (Section \ref{sec:einstein}),
which is approximately equivalent to the ill-defined IAU predecessor
to TCB, known as TDB (see \citealt{sta98c}). The continued use of
$T_{\rm eph}$ or TDB as a barycentric coordinate time for pulsar
timing is deprecated, since under this condition the scale of
units of fitted parameters involving length and/or time differ subtly
from the standard SI definition and fail to comply with IAU Resolutions:
see Paper I.

\subsection{Earth orientation}
In its calculation of the observatory position in the barycentric
frame, \tempo\ uses the IAU 1976 precession \citep{llfm77} and IAU
1980 nutation \citep{sei82a} theories, which are in error at the 50
mas level. More significantly, it also neglects polar motion, which
amounts to up to $\pm 300$ mas, corresponding to up to $\pm 35$ ns of
timing error, with a variable diurnal signature. The polar motion
error is dominated by two periodicities, one with a period of one year
(due to seasonal influences on oceans and the atmosphere), and the
other with a period of 435 days (the Chandler wobble, due to free
precession)-- see Paper I. Depending on the site position, source
direction and hour angle distribution, part of the annual polar motion
may enter the timing model as an annual term at the level of up to
$\sim 10$ ns, corrupting the fitted pulsar position at the level of a
few microarcsec.

\subsection{Shapiro delays}
\tempo\ includes only the first-order Shapiro delay term due to the Sun.
Terms of up to 180 ns for the planetary Shapiro delays, and 9 ns for
the second-order term of the Solar Shapiro delay were omitted.

\subsection{Dispersion}
\tempo\ includes some small errors in the transformation of the
observing frequency to the barycentric reference frame. Firstly, the
computation of the site velocity (equation \ref{eq:roemer_rate})
neglected polar motion, precession and nutation.  Polar motion changes
the direction of the velocity vector $\vec{\dot{x}}$ by up to 300 mas,
altering its projection on the line of sight by up to $\sim 10^{-10}
c$ and corrupting the dispersion term $\Delta_{\rm IS}$ of equation
(\ref{eq:dm}) by a fractional amount $\sim 10^{-9.5}$.  More significantly,
precession moves the vector by $\sim$20 arcsec yr$^{-1}$, leading to a
fractional error in $\Delta_{\rm IS}$ with an amplitude increasing at
a rate of up to $\sim 2\times 10^{-8}$ yr$^{-1}$. This will manifest as
an apparent annual variation in the dispersion measure, with an amplitude
equal to $\sim\mathrm{DM}\cdot 10^{-8} t/$yr, where $t$ is the time
elapsed since Julian year 2000.0. For single-frequency datasets, the
effect would be absorbed by the annual proper motion term, corrupting it
at the $\sim 10 \mu$as yr$^{-1}$ level.

%

Secondly, \tempo\ neglected to account for the Einstein rate
(Section \ref{sec:isdm}). The use by \tempo\ of TDB instead of TCG
(Section \ref{sec:tempo_einstein}) removes a large positive mean from
the Einstein rate, leaving residual rate errors at the $10^{-9.5}$
level, mostly modulated with an annual periodicity.  This corresponds
to fractional errors in $\Delta_{\rm IS}$ at the $10^{-9}$ level.

\subsection{Secular and binary motion}
The full geometric time delay of equation (\ref{eq:geometric})
contains several terms involving the displacement of the pulsar due to
secular motion ($\vec{k}$). Of these, \tempo\ includes only those
terms related both to secular and annual motion. In doing so, it
neglected direct Doppler shift (which includes both classical and
relativistic terms) due to radial velocity and its change in time due
to radial acceleration, and the Shklovskii effect and its change over
time due to the radial motion and transverse acceleration. None of
these omissions is necessarily problematic, because all can be absorbed in a
redefinition of other measured parameters. In \tempoII\ they are
provided directly so that advantage may be taken of additional
constraints that may be available: for example, if the intrinsic spin
orbital period derivative is known to be small, it can be held fixed at
zero and the distance determined directly via the Shklovskii effect.

The public distribution of \tempo\ also omits the Kopeikin terms, however
the secular changes to the viewing geometry may be absorbed in other
parameters, while modified versions of \tempo\ exist that include the
annual-orbital parallax. The orbital parallax, not modelled in any known
version of \tempo, contributes a timing term of up to 30 ns in amplitude.

\section{Summary and Conclusions}
We have presented a model for pulsar pulse times of arrival that
exceeds all previous efforts in its accuracy. This timing model is the
basis of \tempoII, a new software package for pulsar timing. The goal
of \tempoII\  to provide accuracy at the 1-nanosecond level is largely
satisfied by the model presented here, the exception being pulsars
that are part of extremely relativistic binary systems. Further theoretical
work on second- and higher-order relativistic effects will be required
before progress can be made in this area. Fortunately, the pulsars for
which the most precise measurements are possible tend not to be part of
such relativistic binaries.

Terrestrial clock instability, Solar system ephemeris errors and the
local gravitational wave background stand as the likely sources of significant
systematic error. Pulsar timing array campaigns should eventually be able
to provide measurements of these effects, allowing for their removal
from the residuals and more importantly providing a valuable insight into
the causative physical processes.  

With the next generation of radio telescopes, such as the Square
Kilometre Array, achievable measurement error will be reduced in some
cases to below 10~ns. For systematic effects, the timing model
presented here should remain applicable in most cases, and will
present a useful starting point for extension if necessary. However,
at this level of precision previously negligible stochastic effects
will come into play. Possibly the most problematic of these will be
the effect of interstellar scattering.

\section*{Acknowledgements} The authors wish to thank M. Kramer for
comments that improved the manuscript, and A. Irwin, J. Weisberg,
P. Wallace and M. Kramer for useful discussions.

\clearpage
\suppressfloats
\appendix
\section{Tables of parameters and variables}

\begin{table*}
\centering
\begin{minipage}{\linewidth}

\caption{Parameters of the timing model}
\begin{tabular}{clll}
\multicolumn{2}{c}{Name} & Description & Section \\
Algebraic & ASCII \\
\hline\\
$\alpha$ & \textsc{ra} & Right ascension of pulsar (ICRS) & \ref{sec:roemer} \\
$\beta$  & \textsc{elat} & Ecliptic latitude of pulsar & \ref{sec:roemer}\\
$\gamma$ & \textsc{gamma} & Amplitude of binary Einstein delay & \ref{sec:binaryeinstein}\\ 
$\Delta\phi_n$ & \textsc{glph\_}$n$ & Pulse phase increment for glitch $n$ & \ref{sec:phase} \\
$\Delta\nu_n$ & \textsc{glf0\_}$n$ & Pulse frequency increment for glitch $n$ & \ref{sec:phase} \\
$\Delta\dot\nu_n$ & \textsc{glf1\_}$n$ & Pulse frequency derivative increment for glitch $n$ & \ref{sec:phase} \\
$\Delta\nu_{{\rm t}n}$ & \textsc{glf0d\_}$n$ & Amplitude of decaying pulse frequency term for glitch $n$ & \ref{sec:phase} \\
$\delta$ & \textsc{dec} & Declination of pulsar (ICRS) & \ref{sec:roemer} \\
$\delta_\theta$ & \textsc{dtheta} & Relativistic longitudinal deformation of orbit & \ref{sec:binarykinematics}\\
$\delta_r$ & \textsc{dr} & Relativistic radial deformation of orbit & \ref{sec:binarykinematics} \\
$\epsilon_0$ & \textsc{eclobl} & Mean obliquity of the ecliptic at J2000.0 & \ref{sec:roemer}\\
$\zeta$ & \textsc{fddi} & Exponent of frequency-dependent delay & \ref{sec:ISM} \\
$\eta_0$ & \textsc{eps1} & First Laplace-Lagrange parameter at $t^{\rm psr}=T_0$ & \ref{sec:ell1}\\
$\dot\eta$ & \textsc{eps1dot} & Time derivative of first Laplace-Lagrange parameter  & \ref{sec:ell1}\\
$\kappa_0$ & \textsc{eps2} & Second Laplace-Lagrange parameter at $t^{\rm psr}=T_0$ & \ref{sec:ell1}\\
$\dot\kappa$ & \textsc{eps2dot} & Time derivative of second Laplace-Lagrange parameter  & \ref{sec:ell1}\\
$\lambda$ & \textsc{elong} & Ecliptic longitude of pulsar & \ref{sec:roemer}\\
$\mu_\alpha$& \textsc{pmra} & Proper motion in right ascension & \ref{sec:roemer}\\
$\mu_{\rm \alpha B}$& \textsc{kpmra} & Proper motion in right ascension for Kopeikin term& \ref{sec:binarykinematics}\\
$\mu_\beta$& \textsc{pmelat} & Proper motion in ecliptic latitude & \ref{sec:roemer}\\
$\mu_{\rm \beta B}$& \textsc{kpmelat} & Proper motion in ecliptic latitude for Kopeikin term & \ref{sec:binarykinematics}\\
$\mu_\delta$& \textsc{pmdec} & Proper motion in declination & \ref{sec:roemer}\\
$\mu_{\rm \delta B}$& \textsc{kpmdec} & Proper motion in declination for Kopeikin term& \ref{sec:binarykinematics}\\
$\mu_\lambda$& \textsc{pmelong} & Proper motion in ecliptic longitude & \ref{sec:roemer}\\
$\mu_{\rm \lambda B}$& \textsc{kpmelong} & Proper motion in ecliptic longitude for Kopeikin term& \ref{sec:binarykinematics}\\
$\mu_\parallel$& \textsc{pmrv} & Radial proper motion & \ref{sec:roemer}\\
$\nu^{(n)}$ & \textsc{f}$n$ & $n$-th time derivative of pulse frequency at $t^{\rm psr}=t_{\rm freq}$ & \ref{sec:phase} \\
$\csc \xi$ & \textsc{spincsci} & Cosecant of inclination angle of spin axis & \ref{sec:binarygr} \\
$\Pi$ & \textsc{px} & Annual parallax & \ref{sec:roemer}\\
$\tau_{{\rm g}n}$ & \textsc{gltd\_}$n$ & Time constant for glitch $n$ & \ref{sec:phase}\\
$\phi_0$ & & Pulse phase at $t^{\rm psr}=t_{\rm freq}$ (implicit floating parameter) & \ref{sec:phase}\\
$\chi$ & \textsc{spinpa} & Position angle of spin axis w.r.t $\Omega$  & \ref{sec:binarygr} \\
$\Omega$ & \textsc{kom} & Position angle of ascending node & \ref{sec:binarykop}, \ref{sec:binarykinematics}\\
$\omega_0$ & \textsc{om} & Longitude of periastron at $t^{\rm psr}=T_0$ & \ref{sec:binarykinematics} \\
$\dot\omega$ & \textsc{omdot} & Time derivative of longitude of periastron  & \ref{sec:binarykinematics} \\
$\dot\omega_x$ & \textsc{xomdot} & Non-GR contribution to derivative of longitude of periastron  & \ref{sec:binarygr} \\
$\ddot\omega$ & \textsc{om2dot} & Second time derivative of longitude of periastron  & \ref{sec:binaryms} 
\end{tabular}
\end{minipage}
\label{tab:paramaeters}
\end{table*}

\begin{table*}
\centering
\begin{minipage}{\linewidth}
\contcaption{}
\begin{tabular}{clll}
\multicolumn{2}{c}{Name} & Description & Section \\
Algebraic & ASCII \\
\hline\\
$A$ & \textsc{a0} & First binary aberration parameter & \ref{sec:binaryaberration} \\
$B$ & \textsc{b0} & Second binary aberration parameter & \ref{sec:binaryaberration} \\
$a_\perp$ & \textsc{racc} & Radial acceleration of BB-SSB & \ref{sec:vpd} \\
$a_\mu$  & \textsc{pmacc} & Acceleration of BB-SSB in direction parallel to proper motion & \ref{sec:vpd}\\
$d_{\rm Shk}$ & \textsc{d\_shk} & Distance used for Shklovskii term & \ref{sec:vpd} \\
$d_{\rm AOP}$ & \textsc{d\_aop} & Distance used for annual-orbital parallax term & \ref{sec:binarykop} \\
$d_{\rm OP}$ & \textsc{d\_op} & Distance used for orbital parallax term & \ref{sec:binarykop} \\
DM$^{(n)}$ & \textsc{dm}$n$ & $n$-th time derivative of interstellar dispersion measure & \ref{sec:isdm}\\
$e_0$ & \textsc{ecc} & Eccentricity of binary orbit at $t^{\rm psr}=T_0$ & \ref{sec:binarykinematics} \\
$\dot e$ & \textsc{eccdot} & Time derivative of eccentricity of binary orbit & \ref{sec:binarykinematics} \\
$i$ & \textsc{ki} & Binary orbital inclination angle used for Kopeikin terms & \ref{sec:binarykop},\ref{sec:binarykinematics} \\
$k_{\rm f}$ & \textsc{fddc} & Scale of frequency-dependent delay & \ref{sec:ISM} \\
$M$ & \textsc{mtot} & Total binary system mass & \ref{sec:binarygr} \\
$m_2$ & \textsc{m2} & Binary companion mass & \ref{sec:binarygr} \\
$n_0$ & \textsc{ne1au} & Mean electron density at 1 AU heliocentric radius & \ref{sec:ssdm} \\
$P_{\rm b0}$ & \textsc{pb} & Binary orbital period at $t^{\rm psr}=T_0$ & \ref{sec:binarykinematics}\\
$\dot P_{\rm b0}$ & \textsc{pbdot} & Time derivative of binary orbital period  & \ref{sec:binarykinematics}\\
$\dot P_{\rm bx}$ & \textsc{xpbdot} & Non-GR contribution to time derivative of binary orbital period  & \ref{sec:binarygr}\\
$r$ & \textsc{r} & Shapiro delay range parameter & \ref{sec:binaryshapiro} \\
$s$ & \textsc{s} & Shapiro delay shape parameter & \ref{sec:binaryshapiro} \\
$\sin i$ & \textsc{sini} & Sine of orbital inclination & \ref{sec:binarygr}\\
$t_{\rm D}$ & \textsc{dmepoch} & Epoch of dispersion measure & \ref{sec:ISM} \\
$t_{{\rm g}n}$ & \textsc{glep\_}$n$ & Epoch of $n$-th glitch & \ref{sec:phase}\\
$t_{\rm pos}$ & \textsc{posepoch}& Epoch of position & \ref{sec:frames},\ref{sec:geometric},\ref{sec:roemer} \\
$t_{\rm freq}$ & \textsc{pepoch} & Epoch of pulse frequency & \ref{sec:phase}\\
$T_0$ & \textsc{t0} & Time of periastron & \ref{sec:binarykinematics} \\
$T_{\rm asc}$ & \textsc{tasc}& Time of ascending node, Laplace-Lagrange parameterisation & \ref{sec:ell1}\\
$v_\parallel$ & \textsc{rv} & Radial velocity for Doppler correction & \ref{sec:vpd},\ref{sec:doppler}\\
$v_{\parallel {\rm Shkdot}}$ & \textsc{rvs} & Radial velocity for secular acceleration of Shklovskii term & \ref{sec:vpd}\\
$v_\perp$ & \textsc{tv} & Transverse velocity for Doppler correction & \ref{sec:doppler}\\
$x$ & \textsc{a1} & Projected semi-major axis of binary orbit & \ref{sec:binarykinematics}\\
$\dot{x}$ & \textsc{a1dot} & Time derivative of projected semi-major axis of binary orbit & \ref{sec:binarykinematics}\\
$\ddot{x}$ & \textsc{x2dot} & Second time derivative of projected semi-major axis of binary orbit & \ref{sec:binaryms}\\
$z_s$ & \textsc{shapmax} & Shapiro delay parameter, $z_s \equiv -\ln(1-s)$ & \ref{sec:binaryshapiro} \\
\end{tabular}
\end{minipage}
\end{table*}

\begin{table*}
\centering
\begin{minipage}{\linewidth}
\caption{Other symbols used}
\begin{tabular}{cll}
Symbol & Description & Section \\
\hline\\ 
$\unitvec{\alpha}$ & Unit vector in direction of increasing right ascension & \ref{sec:roemer} \\
$\unitvec{\beta}$ & Unit vector in direction of increasing ecliptic latitude & \ref{sec:roemer}\\ 
$\Gamma$ & Doppler shift due to BB--SSB velocity & \ref{sec:doppler}\\
$\Delta_\odot$ & Solar system part of timing formula & \ref{sec:delta_SSB}\\
$\Delta_{\rm A}$ & Atmospheric propagation delay & \ref{sec:atm}\\
$\Delta_{\rm AB}$ & Binary aberration delay & \ref{sec:binaryaberration}\\
$\Delta_{\rm B}$ & Binary part of timing formula & \ref{sec:binary}\\
$\Delta_{{\rm D}\odot}$ & Solar system dispersion delay & \ref{sec:ssdm} \\
$\Delta_{{\rm E}\odot}$ & Solar system Einstein delay & \ref{sec:einstein}\\
$\Delta_{{\rm E}\odot-\oplus}$ & Solar system Einstein delay for geocentre & \ref{sec:einstein}\\
$\Delta_{\rm EB}$ & Binary Einstein delay & \ref{sec:binaryeinstein}\\
$\Delta_{\rm ES}$ & Einstein delay due to secular motion & \ref{sec:doppler}\\
$\Delta_{\rm FDD}$ & Frequency-dependent delay & \ref{sec:isdm}\\
$\Delta_{\rm hz}$ & Atmospheric hydrostatic delay for zenith & \ref{sec:atm} \\
$\Delta_{\rm p}$& Annual parallax delay term & \ref{sec:roemer} \\
$\Delta_{\rm ISD}$ & Interstellar dispersion delay & \ref{sec:isdm}\\
$\Delta_{\rm IS}$ & Interstellar part of timing formula & \ref{sec:ISM} \\
$\Delta_{\rm KB}$ & Kopeikin delay & \ref{sec:binarykop}\\
$\Delta_{\rm R\odot}$ & Solar system Roemer delay & \ref{sec:roemer}\\
$\Delta_{\rm RB}$ & Binary Roemer delay & \ref{sec:binaryroemer}\\
$\Delta_{\rm RB\parallel}$ & Radial part of binary Roemer delay & \ref{sec:binaryroemer}\\
$\Delta_{\rm RE}$ & Combined binary Roemer and Einstein delay & \ref{sec:oldbinary}\\
$\Delta_{\rm S\odot}$ & Solar system Shapiro delay & \ref{sec:shapiro}\\
$\Delta_{\rm S\odot2} $ & Second-order Solar Shapiro delay & \ref{sec:shapiro}\\$\Delta_{\rm SB}$ & Binary Shapiro delay & \ref{sec:binaryshapiro}\\
$\Delta_{\rm VP}$ & Interstellar vacuum propagation delay & \ref{sec:vpd}\\
$\Delta_{wz}$ & Atmospheric wet delay for zenith & \ref{sec:atm}\\
$\Delta L_C^{\rm (PN)}$ & Higher-order post-Newtonian correction to Einstein delay & \ref{sec:einstein} \\
$\Delta L_C^{\rm (A)}$ & Asteroid correction to Einstein delay & \ref{sec:einstein}\\
$\unitvec{\delta}$ & Unit vector in direction of increasing declination & \ref{sec:roemer}\\ 
$\eta$ & First Laplace-Lagrange parameter & \ref{sec:ell1}\\
$\Theta$ & Source elevation angle & \ref{sec:atm}\\
$\theta$ & Orbital longitude & \ref{sec:binarykinematics}\\
$\kappa$  & Second Laplace-Lagrange parameter  & \ref{sec:ell1}\\
$\unitvec{\lambda}$ & Unit vector in direction of increasing ecliptic longitude & \ref{sec:roemer}\\ 
$\vec{\mu}$ & Three-dimensional proper motion vector & \ref{sec:roemer}\\
$\rho$ & Pulsar-Sun-observatory angle & \ref{sec:ssdm} \\ 
$\phi$ & Pulse phase & \ref{sec:phase}\\
$\phi_{{\rm g}n}$ & Contribution to pulse phase by glitch $n$ & \ref{sec:phase}\\
$\varphi$ & Geodetic latitude of the observatory & \ref{sec:atm}\\
$\Phi$ & Orbital phase-type variable for Laplace-Lagrange binary parameterisation & \ref{sec:atm}\\
$\Phi^0$ & Zeroth-order approximation to $\Phi$ & \ref{sec:oldbinary}\\
$\psi_i$ & Pulsar-telescope-object angle for $i$-th solar system body & \ref{sec:shapiro}\\
$\omega$  & Longitude of periastron  & \ref{sec:binarykinematics} \\
$\vec{\omega_\oplus}$  & Angular velocity of Earth rotation  & \ref{sec:roemer} 
\end{tabular}
\end{minipage}
\label{tab:variables}
\end{table*}
\begin{table*}
\centering
\begin{minipage}{\linewidth}
\contcaption{}
\begin{tabular}{cll}
Symbol & Description & Section \\
\hline\\ 
$A_e(u)$ & True anomaly of orbit as a function of eccentric anomaly & \ref{sec:binarykinematics}\\
$\vec{a}$ & Acceleration vector of BB - SSB & \ref{sec:geometric},\ref{sec:vpd}\\
$a$      & Semi-major axis of pulsar orbit & \ref{sec:binarykinematics} \\

$\vec{b}$      & Vector from BB to pulsar & \ref{sec:geometric}, \ref{sec:binarykinematics} \\
$c$      & Vacuum speed of light $\equiv 2.99792458\times 10^{8}$ m s$^{-1}$\\
$C$      & Convenience variable in binary motion & \ref{sec:binarykinematics}\\
$d_p$    & Parallax distance $\equiv  1\;{\rm AU}/\Pi$ & \ref{sec:roemer}\\
$e$,$e_\theta$, $e_r$ & Orbital eccentricities & \ref{sec:binarykinematics} \\
$\unitvec{e_1},\unitvec{e_2}$ & Basis vectors in plane of sky (i.e. $\unitvec{\alpha},\unitvec{\delta}$ or $\unitvec{\lambda},\unitvec{\beta}$) & \ref{sec:binarykinematics} \\
$D$      & Interstellar dispersion constant & \ref{sec:isdm} \\
DM & Interstellar dispersion measure & \ref{sec:isdm} \\
$\mat{E}$ & Matrix relating equatorial and ecliptic coordinates & \ref{sec:roemer}\\
$f$  & Frequency of electromagnetic wave & \ref{sec:ssdm}, \ref{sec:isdm} \\
$f^{\rm SSB}$  & Frequency of electromagnetic wave in SSB frame & \ref{sec:ssdm}, \ref{sec:isdm} \\
$G$  & Gravitational constant \\
$H$  & Height of observatory above the geoid & \ref{sec:atm}\\
$i$  & Orbital inclination angle & \ref{sec:binarykinematics} \\
$\vec{k}$ & Displacement of BB since $t^{\rm SSB} = t_{\rm pos}$ & \ref{sec:geometric}, \ref{sec:roemer}, \ref{sec:vpd}\\
$k$ & Derivative of longitude of periastron w.r.t. true anomaly & \ref{sec:binarykinematics} \\
$k_{\rm D}$ & Constant of proportionality relating $D$ and DM & \ref{sec:isdm} \\
$m_2$  & Binary companion mass & \ref{sec:binaryshapiro} \\
$m_h(\Theta)$ & Mapping function for hydrostatic propagation delay & \ref{sec:atm}\\
$m_w(\Theta)$ & Mapping function for wet propagation delay & \ref{sec:atm}\\
$n$ & Mean angular frequency of binary since $T_0$ & \ref{sec:binarykinematics}\\
$P$ & Surface atmospheric pressure & \ref{sec:atm}\\
$P_{\rm b}$ & Binary orbital period & \ref{sec:binarykinematics}\\
$\mat{Q}$ & Frame bias, precession and nutation matrix & \ref{sec:roemer}\\
$\mat{R}$ & Earth rotation matrix & \ref{sec:roemer}\\
$\vec{r}$ & BCRS position vector of observatory & \ref{sec:geometric}, \ref{sec:roemer}\\
$\vec{r_\oplus}$ & BCRS position vector of geocentre & \ref{sec:einstein}, \ref{sec:roemer}\\
$\vec{R}$ & BCRS displacement vector of pulsar minus observatory & \ref{sec:geometric} \\
$\vec{R_0}$ & BCRS position vector of binary barycentre at $t^{\rm SSB} = t_{\rm pos}$ &  \ref{sec:geometric} \\
$\vec{R}_{\rm BB}$ & BCRS position vector of BB at time of observation & \ref{sec:roemer} \\
$\unitvec{R}_{\rm BB}$ & BCRS direction unit vector of BB at time of observation & \ref{sec:roemer} \\
$\vec{s}$ & GCRS position vector of observatory & \ref{sec:einstein}, \ref{sec:roemer}\\
$S$      & Convenience variable in binary motion & \ref{sec:binarykinematics}\\
$s_{\rm IS}$ & ISM part of geometric path length (minus value at $t^{\rm SSB} = t_{\rm pos}$) & \ref{sec:vpd}\\
$s_{\rm B}$ & Binary orbital part of geometric path length & \ref{sec:binaryroemer}\\
$t_0$ & Common epoch of TT, TCG and TCB & \ref{sec:einstein}\\
$t^{\rm obs}$& Terrestrial Time ($\sim$ proper time) at observatory & \ref{sec:frames}\\
$t^{\rm SSB}$& Barycentric Coordinate Time & \ref{sec:frames} \\
$t^{\rm BB}$& Coordinate time in frame of binary barycentre & \ref{sec:frames}\\
$t^{\rm psr}$& Proper time in pulsar frame & \ref{sec:frames} \\
$t^{\rm psr}_{\rm e}$& Proper time of emission of pulse & \ref{sec:frames}\\
$t^{\rm frame}_{\rm a}$ & Time of arrival of pulse in a given frame & \ref{sec:frames} \\
$T_\odot$ & $GM_\odot/{c^3}$, half the light travel time across the Solar Schwarzschild radius & \ref{sec:binaryshapiro} \\
$U_\oplus$& Gravitational potential at the geocentre due to SS bodies & \ref{sec:einstein}\\
$u$ & Eccentric anomaly of orbit & \ref{sec:binarykinematics} \\
$v_\oplus$& Velocity of the geocentre relative to the SSB & \ref{sec:einstein}\\
$\mat{W}$ & Polar motion matrix & \ref{sec:roemer}\\
$W_0$ & Gravitational plus spin potential of Earth at the geoid & \ref{sec:einstein}\\

\end{tabular}
\end{minipage}
\end{table*}

\clearpage

\bibliographystyle{mn2e}
\bibliography{journals,modrefs,psrrefs,crossrefs,tempo2}
\end{document}